\documentclass[final,3p,times]{elsarticle}

\usepackage{amssymb}

\usepackage{lineno}

\usepackage[T1]{fontenc}

\usepackage{setspace}

\usepackage{longtable}

\usepackage{natbib}

\journal{a Software Engineering Journal}

\begin{document}

\begin{frontmatter}



\title{ABCDE \textendash Agile Block Chain Dapp Engineering}


\author{Lodovica Marchesi, Michele Marchesi, Roberto Tonelli}

\address{DMI, University of Cagliari}
\ead{lodo.marchesi@gmail.com, marchesi@unica.it, roberto.tonelli@dsf.unica.it}

\begin{abstract}
Cryptocurrencies and their foundation technology, the Blockchain, are reshaping finance and economics, allowing a decentralized approach enabling trusted applications with no trusted counterpart. 
More recently, the Blockchain and the programs running on it, called Smart Contracts, are also finding more and more applications in all fields requiring trust and sound certifications. 
Some people have come to the point of saying that the ``Blockchain revolution'' can be compared to that of the Internet and the Web in their early days. 
As a result, all software development revolving around the Blockchain technology is growing at a staggering rate. The feeling of many software engineers about such huge interest in Blockchain technologies is that of unruled and hurried software development, a sort of competition on a first-come-first-served basis which does not assure neither software quality, nor that the basic concepts of software engineering are taken into account.
This paper tries to cope with this issue, proposing a software development process to gather the requirement, analyze, design, develop, test and deploy Blockchain applications. The process is based on several Agile practices, such as User Stories and iterative and incremental development based on them. However, it makes also use of more formal notations, such as some UML diagrams describing the design of the system, with additions to represent specific concepts found in Blockchain development. The method is described in good detail, and an example is given to show how it works.
\end{abstract}


\end{frontmatter}


\section{Introduction}
\label{S:1}

The so-called "decentralized applications", or "dApps", are one of the main new trends of software development. 
dApps typically run on a blockchain, the technology originally introduced to manage the Bitcoin digital currency \cite{nakamoto2008}. 
Blockchain software runs in a network of peer-to-peer nodes, so it is naturally decentralized, redundant and transparent. 
A few years after the introduction of Bitcoin in 2009, developers and managers realized that a blockchain can be also the ideal environment for a decentralized computer. 
This led to the introduction of Ethereum blockchain, a network whose nodes are also able to run Turing-complete programs \cite{wood2014}, called "Smart Contracts" (SCs) following and idea of Nick Szabo \cite{szabo1997}.
SCs are general computer programs, though with some specific features. 
The main idea behind them is that they can be used for the automated enforcement of contractual obligations, without having to trust a central authority, and without space and time constraints. 
So, there is a huge wave of interest in SCs and applications of the blockchain, especially in the financial realm. 
Some authors even said that \textit{"we should think about the Blockchain as another class of thing like the Internet [...]"} \cite{swan2015} and that the \textit{"wide adoption of Blockchain technology has the potential of reshaping the current financial services technical infrastructure."} \cite{biella2016}.

This interest led to a huge amount of money flooding into Blockchain ventures. 
During 2017, this steady inflow of money, paired with the limited amount of available digital money -- a feature that most digital currencies have by design -- made the price of Bitcoin and other digital currencies spike at the end of the year. 
This peak was mainly ignited by the Initial Coin Offers (ICO) phenomenon, where a startup publishes a white paper describing their idea, and gathers digital money by issuing a token, which is a second level digital currency, managed by a SC \cite{fenu2018}. 
This token can then be immediately traded on an exchange (a pseudo bank running on the Web, which allows to exchange digital currencies against traditional money, or against other digital currencies).
The enthusiasm for this new idea, the ever increasing prices and profits, and the \textit{fear of missing out} (FOMO) led to many billions USD 
pouring into tokens. 
At the beginning of 2018 the bubble deflated, with the global capitalization of digital currencies going from more than 800 billion USD on January 7, 2018, to about 115 billion USD on February 2019.
During 2019, however, a renewed interest to digital currencies lead their global market cap back to about 200 billion USD as of September.
Huge inflows of money and venture capital were also poured into blockchain initiatives not linked to digital currencies. 
These are the so called "permissioned" blockchains, or distributed ledgers (DL), intended to be run by a set of nodes chosen by invitation.

All the initiatives behind blockchain technology -- new digital currencies with their own blockchain, exchanges and other Web-based ventures using digital money, ICO startups, applications running on permissioned blockchains or DLs -- are based on developing a new software system. 
This often led to a run to be the first on the market, as always happens with new technology waves, with quick application development, neglecting good development practices, and often even basic testing and security assessment. 

Some big disasters quickly followed, with a total of literally billions of USD (at least at the nominal exchange rate) of digital currencies stolen or lost.
Several exchanges were hacked, since the beginning of Bitcoin trading, for a total of various billion equivalent USD (for a list up to 2017, see  \cite{chohan2018}). 
Also, SCs were often exploited, taking advantage of their novelty and of the hurried software development \cite{atzei2017}, \cite{destefanis2018}.

Overall, the scenario looks that of a competition on a first-come-first-served basis, where the basic principles and practices of software engineering (SE) are not taken into account. 
The quality of the resulting software is accordingly compromised.

It is well known that, to develop a reliable and maintainable software system, one needs to follow an explicit development process, and use sound SE practices.
Among the latters, in the context of blockchain development, we stress the importance of requirement elicitation, system design, specific notations and security assessment.
In essence, we need blockchain-oriented software engineering (BOSE) \cite{porru2017}.

In this paper we present a development process for applications based on Smart Contracts running on a blockchain, which are usually called "dApps" (decentralized applications).
The process covers all the standard phases of software life cycle: requirement elicitation, design, implementation, security assessment and testing, and ongoing maintenance.
We call the process "ABCDE", Agile Block Chain Dapp Engineering.
ABCDE is an agile software development process, meaning that is follows the principles of Agile Manifesto \cite{beck2001}. 
However, we had to complement the agile process with a more formal approach, using UML diagrams with a specific notation for SCs, and a specific checklist for security assessment. 

The first question we had to answer regarding ABCDE is "why a new process"? 
Why not use an existing process, waterfall or agile, for dApp development? 
The answer to this question stems from the observation that a SC is very peculiar software. 
It runs on all the nodes hosting a Blockchain, and its execution has the strong constraint that all outputs and state changes resulting from SC execution must be the same in all nodes. 
Consequently, a SC is strictly forbidden to access the external word -- it can answer to external messages belonging to its public interface, and can send messages to other SCs running on the same blockchain; no other kind of interaction with the external world is allowed. 
This fact implies that any dApp is intrinsically divided in two subsystems -- the SCs running on a blockchain, and the applications allowing users and devices to interact with the SCs. 
Another specificity of SC realm is the need to introduce new concepts with respect to traditional programming, like those of "address" to refer to SCs; of signed "transaction" to send a message starting from a given address; of "GAS" needed to run a SC; of digital money owned by, and transferred between, SCs; of "oracle", a SC able to provide data coming from the external world without violating the constraint described above. 
Moreover, referring to Solidity, the programming language of Ethereum, which is presently the most used blockchain actually running SCs, there are further specific concepts, such as those of "modifier" (a boolean function acting as a guard to the execution of another function), of "library contract", and there are constraints on the use of typical structures of object oriented programming languages. For instance, inheritance is limited, and the available collections are just arrays and mappings. 
Finally, two papers by Chakraborty, Bosu et al. \cite{chakraborty2018} \cite{bosu2019}, present the results of a survey among blockchain developers. They found that the prevalent opinion is that blockchain development is different from traditional one, due to the strict and non-conventional security and reliability requirements, and to other unique characteristics of the dApp development domain, such as immutability, difficulty in upgrading the software, and so on. More information on this survey is presented in Sec. \ref{S:3}.

This specificity led us to conclude that a new method is needed for dApp development. In fact, ABCDE is not entirely new, but it is a significant extension to classical agile methods, such as Scrum \cite{schwaber2001}. 
With respect to Scrum, ABCDE does not only describe how the development should be managed, but also introduces specific practices such as the use of modified UML diagrams to describe SCs, and checklists for security assessment. 
Other agile practices, such as simplicity, test driven development \cite{beck2002}, refactoring \cite{fowler2018}, collective code ownership, pair programming, are encouraged if the team feels they are useful, but are not prescribed by ABCDE.

The main characteristic of ABCDE is the split of dApp development in two flows which are carried on concurrently, after a common start.
The first flow is the specification and development of the SCs.
The second flow is about the development of the software applications which allow external actors to interact with the SCs. 

In this paper, we also introduce a notation augmenting some UML diagrams (Use Case, Sequence, and Class diagrams) to account for SC specificity in the context of Solidity language. 
The dApp design when other languages to develop SCs are used can be represented with different UML extensions. In this paper, we limit ourselves to Solidity. 

The proposed ABCDE methods has been tested on some real dApp development projects, carried on at our department, and at some firms we are consultant of.

The remainder of this paper is organized as follows. 
In Section \ref{S:2} we describe the architecture of a dApp, and introduce the specific issues and practices needed for dApp development.
In Section \ref{S:3} we present the related work in the same, or similar fields. 
Section \ref{S:4} describes the proposed ABCDE process in every detail, including the modifications of some UML diagrams to cope with Solidity concepts, security assessment and what is needed to extend the notation to other languages for SC development. 
A simplified example, drawn from a real case, is presented in Section \ref{S:5}, together with reporting on actual uses of ABCDE. 
Finally, Section \ref{S:6} presents the conclusions and future work ideas.

\section{Background}
\label{S:2}

\subsection{Decentralized applications} \label{subs:dapp}

We define as dApp (decentralized application) a software system that uses distributed ledger technology (DLT), typically a blockchain\footnote{From now on, we'll use the terms "DLT" and "blockchain" interchangeably.}, as a central hub to store and exchange information, through Smart Contracts (SCs).
Note that it is not a blockchain software able to manage a new cryptocurrency or other applications -- that is, software enabling blockchain nodes, which needs different kinds of of development practices, not the subject of this work.

A blockchain is a distributed data structure -- managed by a set of connected nodes -- characterized by the following elements:

\begin{itemize}
\item it is redundant (each node holds a copy of the blockchain);
\item it is append-only -- once written, the information cannot be changed or deleted;
\item the blockchain state is changed by sending \textit{transactions} to the network -- in \textit{public blockchains}, everyone can send a transaction;
\item all transactions are checked by the reached nodes; invalid ones are ignored;
\item the valid transactions are typically recorded in sequentially ordered blocks -- hence the name "blockchain" -- whose creation is managed by a consensus algorithm among the nodes;
\item all transactions are sent from an unique address, which is in turn computed from a public key. Only the owner of the private key associated with it can sign the transactions coming from this address using asymmetric cryptography, validating them;
\item if the blockchain is able to execute SCs, a transaction can create a SC, or execute one of its public functions; in this case, the function is executed by all nodes, when the transaction is evaluated -- the execution of the program \textit{is} the execution of the transaction.
\end{itemize}

A dApp is usually composed of SCs deployed on a blockchain, and of software able to create and send transactions to them. 
This software usually provides a user interface, running on a PC, or on a mobile device.
Additional information could be stored on a server, and further business logic could be executed on this.

Most present real applications of dApps and SCs are intended for the management of digital currencies or tokens, that have a true monetary value. 
The use of dApps has been introduced also for other scopes, like notarization of information, identity management, voting, games and betting, goods provenance certification, and many others \cite{zheng2018}.

\begin{figure}[ht]
\centering 
\includegraphics[width=9cm]{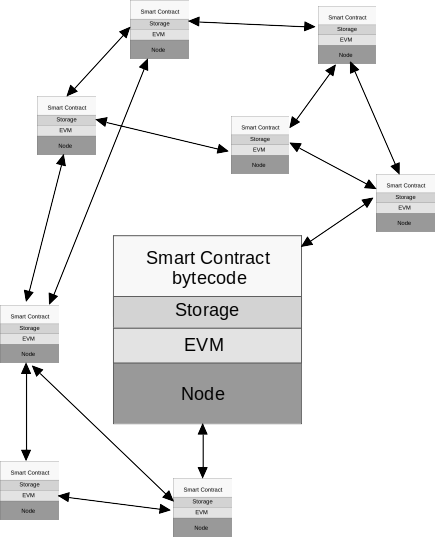}
\caption{The Ethereum Blockchain running a SC. The same SC bytecode is executed by each node.}
\label{fig:fig1}
\end{figure}

In this paper, we will use as a reference Ethereum, whose SCs were the first to exhibit a Turing-complete capability, and is presently the most used blockchain to develop SCs \cite{tikhomirov2017ethereum}. 
Just to quote some figures, as of October 1, 2019 there were 1533 out of 1720 active tokens\footnote{We define as "active" a token whose market cap is over 100,000 USD.} managed by Ethereum SCs, worth more than 8 billion USD at current prices \cite{cmc2019}. 
Moreover, as of November 2019, 2654 dApps were running on Ethereum, out of a total of 3169 surveyed dApps \cite{sod2019}. These figures are about public blockchains. 
Data on dApps running on permissioned blockchains are more difficult to find, but Ethereum is very popular also for this kind of dApps. Open source DLTs such as Hyperledger and Corda are also widely used.

The Ethereum Virtual Machine (EVM), able to execute SC Ethereum bytecode, runs on all nodes of the Ethereum blockchain \cite{dannen2017introducing}.
In practice, the SCs are written in high-level languages (HLL).
Nowadays, the most popular HLL for Ethereum is called Solidity. Other languages, such as Flint and Vyper can be used, but their adoption is still far behind Solidity. 
Fig. \ref{fig:fig1} shows a sketch of the Ethereum blockchain, with the architecture of its nodes running a SC. The original Ethereum software running the node is written in C++ or Go language. 
A compatible implementation written in Rust language (Parity) is also available.

As written in the Introduction, SCs run in an isolated environment. 
The results of their execution must be the same whatever node they run in; consequently, they cannot get information from the external world (which mutates with time), and cannot initiate a computation autonomously (for instance at given times).
SCs can only access and change their state, and send messages to other SCs. 
The state of a SC is permanently stored in the blockchain, using \textit{storage} variables. 
Another SC specificity is their immutability.
Once a SC is deployed, it is in the blockchain forever -- it cannot be modified or erased, though it can be forever disabled.

SCs are created by special transactions.
Creating a SC and changing its state costs units of "GAS", which must be paid in Ether (the digital currency of the Ethereum Blockchain).
Each SC has a unique Ethereum address, that is used to send messages to it.
In Solidity, a SC can inherit from other SCs; it has a public interface, that is a set of functions that can be called through a transaction.
The call of a public function of a SC is called a "message".
Sending a message to a SC can be performed either by posting a transaction coming from an address, or by executing code of the same, or of another SC. 
In the former case, the transaction must be accepted by the network, and it will take time, and a bigger amount of GAS.
In the latter case, the execution is immediate, and the cost is lower.

A SC can receive and send Ethers, from and to another SC, or an address.
A function belonging to a SC can change its state, can call functions belonging to other SCs, including itself, and can create and send a transaction to an address or to another SC.
In the latter case, the transaction is executed immediately.
A function which returns a value without changing the state of its SC or sending a transaction, is executed immediately by the EVM and costs nothing.
This kind of function is said of type "\textit{view}".

\begin{figure}[ht]
\centering 
\includegraphics[width=14cm]{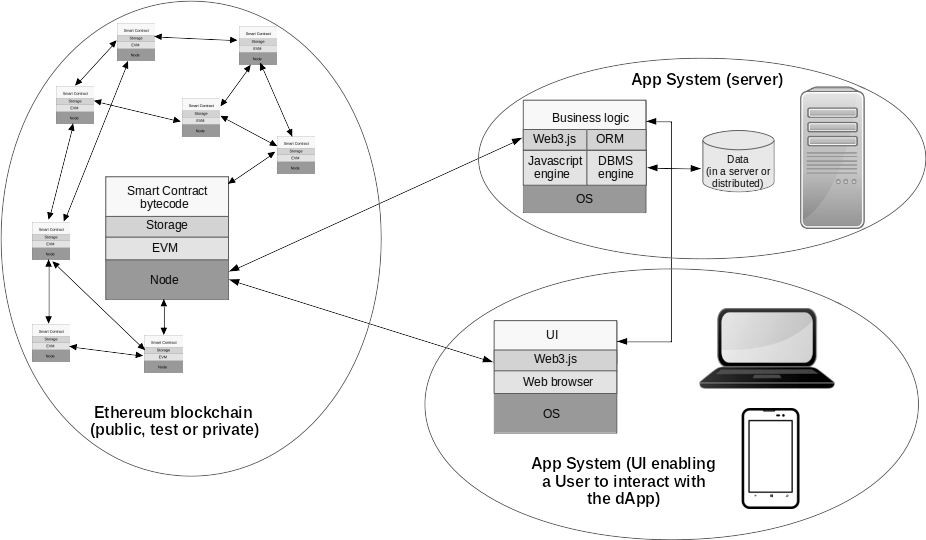}
\caption{A typical architecture of an Ethereum dApp application. The App System is shown on the right, the blockchain with its SCs on the left.}
\label{fig:fig2}
\end{figure}

A typical dApp architecture is shown in Fig. \ref{fig:fig2}.
It is composed of a software system running on mobile devices and/or on servers, possibly on the Cloud, exchanging information with users and external devices, which we call "App System".
Its User Interface (UI) typically runs on a Web browser. It can have a server component, to store data that cannot be stored in the blockchain, and to perform business computations. 
In Ethereum, the App System typically communicates with the blockchain using the "web3.js" Javascript library, which manages the creation and dispatch of transactions.

The other component are the SCs running on the blockchain. In a not trivial system, it is composed of various SCs deployed on the blockhain and identified by their Ethereum address.

\subsection{Agility and dApp Development} \label{subs:agile}

Nowadays, the developments of dApps worldwide share some common characteristics.
Several teams involved are typically working on ICO projects, which gathered money through tokens and are about applications of blockchain technology.
Other projects are promoted by startups trying to take advantage of the novelty of dApps to develop disruptive solutions, or to get a niche where to thrive.
In both cases, they are typically small, self-organizing co-located teams, where experts of system requirements are highly available.

Other characteristics of dApp development is that dApps typically are not life-critical applications, though some of them can be mission-critical.
However, the time-to-market and the ability to get an early feedback from the users and the stakeholders are essential, though often the requirements of the dApp initially are only vaguely defined and are subject to change.

All these features make dApp development an ideal candidate for tha use of Agile Methods (AMs). 
In fact, AMs are suited for small, self-organizing teams, possibly co-located, working on projects whose requirements can change \cite{beck2001}.
AMs are considered to be able to deliver quickly and often, as needed by dApp projects.

The most used AM is presently Scrum, which is iterative and incremental, with short iterations (1-4 weeks) \cite{schwaber2001}. 
Scrum does not prescribe specific software development practices, but is focused on the process.
In short, Scrum, as most other AMs, typically performs requirement elicitation through User Stories (USs), that are short descriptions of how the system answers to inputs from users, or from external devices \cite{cohn2004}. 
USs are mostly gathered at the beginning of the development, but can be modified and augmented at any time. 
The project advances iteratively implementing a subset of the USs at each iteration.
The person in charge of choosing this subset, and explaining their USs to the team is the Product Owner. 

Other agile practices that are well suited to dApp development, and that can be used in the proposed process if the team chooses them are: 

\begin{itemize}

\item \textit{Test Driven Design}: this practice prescribes writing the tests before the code \cite{janzen2005}, using an automated test suite that can be run whenever needed. For the App System, this is the preferred technique, because it guarantees that the Unit Tests are always present, and their development is not indefinitely postponed if the team is under pressure. For SCs written in Solidity, at the moment the most popular testing environment is Truffle \cite{tru2019}. 
\item \textit{Continuous Integration}: the practice of merging all developer working copies to a shared mainline, even several times a day. Developing dApps, this practice is critical, and it should be practiced both on the App System and the SCs, checking at each merging also how the two systems interact through transactions. This practice requires a development environment provided of a working test blockchain, possibly simulated, to deploy SCs and to test all interactions. 
\item \textit{Collective code ownership}: this practice allows every developer to intervene on whatever code s/he considers appropriate to modify. With small, dynamic teams as typically happens with dApp development, this practice should clearly be applied. However, often the team members expert in SC development differ from those expert in App System, so their spheres of influence remain separate.
\item \textit{Refactoring}: this is the attitude to intervene on the code whenever and wherever it can be improved, improving its design without introducing new features. This practice needs to have an automated test suite, that can be run when the refactoring is made, to assess the absence of unwanted side effects. This is especially needed with the complex architecture of dApps, whose components interacts through transactions.
\item \textit{Information Radiators (Cards, Boards, Burndown charts)}: making visible the status of a project using boards that can be observed by everyone and updated in real time, is an practice that is common to all agile projects, and that can obviously greatly benefit also dApp development.
\item \textit{Coding Standards}: the practice of strictly following the same coding standard throughout the code, with proper differentiation between App System code and SCs, should be applied to all projects developed following sound SE practices. However, the dynamicity of the teams and the push to quickly develop applications make necessary that the project manager (or the Scrum Master) ensures that this practice is strictly followed.
\item \textit{Pair Programming (PP)}: this practice is strictly enforced in "pure" Extreme Programming teams \cite{beck2000}. In our approach, we suggest to use PP in the case the software to be developed is critical, is not yet well understood, or there are new team members to train on the job.
\end{itemize}

\subsection{Security Assessment} \label{subs:assur}
In the previous section, we made the case for using agile practices for developing dApps. However, many dApps deal with direct digital currency or token usage, that is with entities that have a direct, real monetary value. 
In other cases, they may deal with contractual issues, again with strong economic implications, as in the case of document certification, supply chain management, voting systems. 
Therefore, in most cases dApps are business-critical, and very strict security requirements should be assured. 
Code inspection, security patterns, and thorough tests must be applied to get a reasonable security level. ABCDE proposed security assessment will be described in detail in Section \ref{subs:security}.

\section{Related Work}
\label{S:3}

SE for dApp development, sometime called Blockchain-Oriented Software Engineering (BOSE) is still in its infancy. 
The first call for BOSE was made in 2017 by Porru et al. 
They highligh "\textit{the need for new professional roles, enhanced security and reliability, novel modeling languages, and specialized metrics}", and propose
"\textit{new directions for blockchain-oriented software engineering, focusing on collaboration among large teams, testing activities, and specialized tools for the creation of smart contracts}" \cite{porru2017}.
They also suggest the adaptation of existing design notations, such as UML, the Unified Modelling Language \cite{rumbaugh2017} to unambiguously specify and document dApps. 

The book by Xu et al. is perhaps the most complete overview of the engineering aspects of blockchains to date \cite{xu2019}. 
Among others, it deals with some SE issues, such as the evaluation of the suitability to use a dApp or not, the selection and configuration of the proper blockchain solution (public, permissioned, private), a collection of patterns for the design of blockchain-based applications, and even model-driven generation of SC code. Some of the topics of the book were introduced previously in \cite{xu2017}.

Wessling et al. propose a method to find how the architecture of an application  could benefit from blockchain technology. 
They identify the actors involved and how they trust each others to derive a high-level hybrid architecture of a blockchain-based application \cite{Wessling2018}.

Fridgen et al. propose an approach for eliciting use cases in the context of blockchain applications, applying action design research method.
Their method is evaluated in four distinct case studies regarding banking, insurance, automotive and construction \cite{fridgen2018}.

Jurgelaitis et al. propose a method based on Model Driven Architecture, which could be used for describing
blockchain-based systems using a general language in order to facilitate blockchain development process \cite{jurg2019}.

A paper by Beller and Hejderup \cite{beller2019} is worth mentioning, though it does not really advocate to use SE practices to develop blockchain applications. Instead, it is about "\textit{how blockchain technology could solve two core SE problems: Continuous Integration (CI) Services such as Travis CI and Package Managers such as apt-get}". The use of SCs to manage agile development, including the automated compensation of developers when their software passes acceptance tests was also proposed by Lenarduzzi et al. \cite{lenarduzzi2018}, \cite{march2019}.

Chakraborty et al. using an online survey got answers from 156 active blockchain software (BCS) developers, finding that "\textit{standard software engineering methods including testing and security best practices need to be adapted with more seriousness to address unique characteristics of blockchain and mitigate potential threat}s" \cite{chakraborty2018}.
The same authors published an extended version of the same research, further highlighting that there is a need for "\textit{an array of new or improved tools, such as: customized IDE for BCS development tasks, debuggers for smart-contracts, testing support, easily deployable simulators, and BCS domain specific design notations}" \cite{bosu2019}.
They found that most BCS developers feel that BCS development is different from traditional one, due to the strict and non-conventional security and reliability requirements, and to other unique characteristics of the dApp development domain (e.g., immutability, difficulty in upgrading the software, operations on a complex, secured, distributed and
decentralized network). As anticipated in the Introduction, these findings confirm the expedience to devise a software engineering process such as ABCDE for BCS development.

Regarding dApp security, many publicly available documents, and scientific papers have been already published. 
Among the most recent ones, the survey of Praitheeshan et al. analyzes the literature about Ethereum SC security, summarizing the main security attacks against SCs, their key vulnerabilities, the security analysis methods and tools~\cite{prait2019}. 
They classify analysis methods in static analysis, dynamic analysis, and formal verification, and discuss the relative pros and cons of these classes, also providing a large bibliography with 160 references. 
Huang et al. deal with SC security in a broader way, considering also Hyperledger security, and performing a survey from a software lifecycle perspective~\cite{huang2019}. 
After a classification of security issues in SCs, both in Ethereum and Hyperledger Fabric, they consider the securities activities according to the various phases of dApp development (design, implementation, testing before deployment, and runtime monitoring), quoting several references and giving practical advice. 
These two papers together include references to virtually all the work which have been published about SC security to date.

Various papers have been published to suggest upgrades of Unified Modeling Language \cite{rumbaugh2017} notation to enable it to better represent specific application fields. 
Baumeister et al. described an extension of UML for Hypermedia design, through the addition of a new Navigational Structure Model and new stereotypes  \cite{baumeister1999}.

Baresi et al. extend and customize UML with web design concepts borrowed from the Hypermedia Design Model. Hypermedia elements are described through appropriate UML stereotypes \cite{baresi2001}. 

Rocha and Ducasse \cite{rocha2018preliminary} study SC design and compare three complementary software engineering models -- Entity-Relationship diagrams, UML and BPMN. To better represent SC concepts, they propose a simple addition to UML Class Diagrams, that is a small "chain" icon in the UML class representing a contract as a notation to more easily identify it as a blockchain artifact.

\section{Proposed Method for dApp Development}
\label{S:4}
\subsection{Overall Process} \label{subs:process}

Our approach, ABCDE, takes into account the substantial difference between developing traditional software (the App System) and developing SCs, and separates the two activities. 
For both developments, ABCDE takes advantage of on an agile approach, because agile methods are suited to develop systems whose requirements are not completely understood since the beginning, or tend to change, as it is the case of dApps.
However, a more formal approach with respect to agile development is also added, to address the security and maintenance issues which are very important in dApp development.

\begin{figure}[ht]
\centering 
\includegraphics[width=11cm]{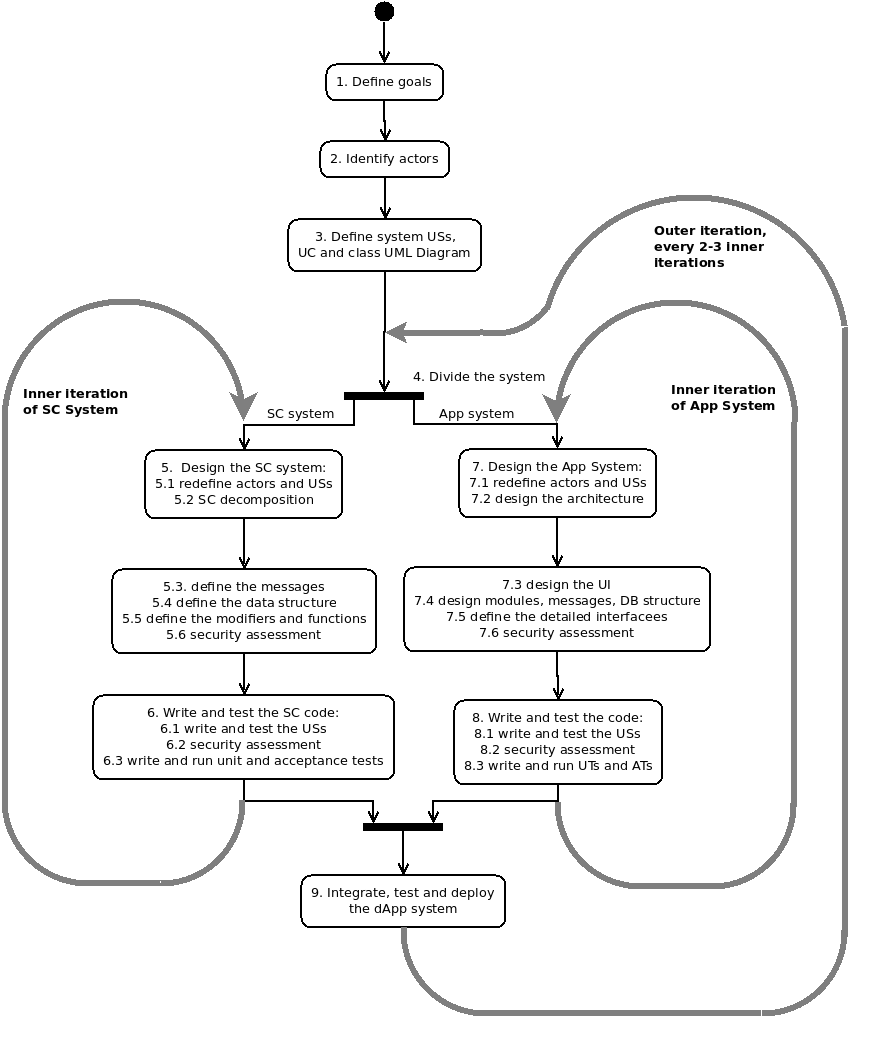}
\caption{The proposed ABCDE process, shown as a UML activity diagram.}
\label{fig:fig3}
\end{figure}

The steps of the proposed ABCDE design method, which is currently focused on Ethereum blockchain and Solidity language, are shown in Fig. \ref{fig:fig3} as UML activity diagram. Note that most steps are in fact performed many times, because the approach is iterative and incremental.

In deeper detail, the proposed development process is the following:

\renewcommand{\labelenumii}{\theenumi.\arabic{enumii}}
\begin{enumerate}
\item \textbf{Goal of the system}. Write 10-30 words summing up the goal, and display them in a place that is visible to the whole team. This is a practice that, as far as we know, was introduced by Coad and Yourdon in their 1991 book on object-oriented analysis \cite{coad1991}, and that we always found useful. It has some similarities with the "Sprint Goal" that Scrum method prescribes to find and make visible to the team, at the beginning of each iteration \cite{schwaber2001}, but here the goal is found for the whole system.
\item \label{item:actors} \textbf{Find the actors.} Identify the actors who will interact with the dApp system.                    
The actors are human roles, and external systems or devices that exchange information with the dApp to build.
\item \label{item:USs} \textbf{User Stories.} The system requirements are expressed as user stories (USs) \cite{cohn2004}, to be able to follow the classical agile approach for project management, used in Extreme Programming \cite{beck2000} and Scrum \cite{schwaber2001}. In this step, the dApp system under development should be considered in full. The decision to develop it using a blockchain, a set of servers, possibly in the cloud, or another architecture, is not important. At this point, we found useful, though not mandatory, to use a UML Use Case Diagram to graphically show the relationships among the actors and the USs. If the decision is taken to implement the system using a blockchain, and related SCs, the following steps are taken.
\item \textbf{Divide the system in two subsystems. }

\begin{enumerate}
\item The SCs running on the blockchain.
\item The App System, that is the external system that interacts with the blockchain, creating and sending transactions, and monitoring the Events that may happen when a SC executes a function.
\end{enumerate}
\item \textbf{Design of the SCs.}

This step is about designing the SCs, using in our case the Solidity language. 
This activity has very peculiar characteristics with respect to standard software design, as highlighted by~\cite{chakraborty2018}.
The activity is not performed in a single step for the whole SC system, but is performed through iterations that include coding and delivering increments of SCs, which are the USs chosen for each iteration. 
Its sub-steps are the following:

\begin{enumerate}
\item Replay steps 2 and 3 (finding Actors and USs) by focusing only on actors directly interacting with the SCs. If external SCs are used by the SCs of the system under development, they should be included among the actors. For each US defined in this step, define also the related acceptance test(s).            
\item  Define broadly the SCs composing the SC subsystem, stating their responsibilities to store information, and the messages they should respond to. For non-trivial systems, you will typically need various interacting SCs. Consider also the use of inheritance for abstracting common features of SCs. Describe in detail the interfaces of the libraries  and of the external SCs used. UML class diagrams with proper additions will be used, as shown later in Sec. \ref{subs:uml}.
\item \label{subitem:flow} Define the flow of messages and Ether transfers among SCs,  external SCs and the App System. Use UML sequence diagrams to document these interactions, if they are non-trivial. If needed, define the state changes of SCs using UML statecharts. 
\item Define in detail the data structure of the SCs, their external interface (Application Binary Interface, ABI) and the relevant events that can be raised by the functions of the SC.
\item Define the internal, private functions and the modifiers -- special functions that are executed before the functions that apply them, and that usually test the preconditions needed before the function can be safely executed.      
\item Define the tests and perform the security assessment practices. This is a very important step because, as already explained above, most SCs are very critical and deal with money. Section \ref{subs:security} in the followings describes in deeper detail the security assessment we use for Ethereum SCs. 
\end{enumerate}

\item \textbf{Coding and testing the SC system.} \label{item:code} 
Following the agile approach, the SC system is built and tested incrementally. 
We advocate following a Scrum approach \cite{schwaber2001}, because it is very effective and popular among developers.
In Scrum, a subset of USs are implemented at each iteration. However, also a Lean-Kanban approach is feasible, implementing the USs in a continuous flow, with the work in progress controlled by the Kanban board \cite{anderson2010}. The coding and testing activities are:
\begin{enumerate}
\item Incrementally write and test the SCs with an agile approach (Scrum or Kanban). Owing to the strict security requirements, typically this activity cannot be performed in a strict incremental way, just implementing one US after another. Instead, starting from the data structure and interfaces of SCs, the overall kernel SC architecture is implemented and tested first. This can be accomplished by using special "User Stories" which are not the description of the interaction with users, but are about the implementation of the architecture of the system. Then, complementary USs can be added.                  
\item Perform the security assessment of the code written for the increment (see section \ref{tab:codCheck}).
\item Write automated Unit Tests (UTs) and Acceptance Tests (ATs) for the SCs and USs implemented, respectively. Add the new tests to the test suite. The most used testing environments for Solidity is Truffle \cite{tru2019}. Run the whole test suite to make sure that the additions did not break the system. 
\end{enumerate}

\item \textbf{Design of the external interaction subsystem (App System).} \label{item:external}

This step is about designing the App System, which interacts with the users and devices, send messages to the blockchain, and can manage its own repositories (data bases and/or documents). 
This activity is very similar to designing a standard Web application. It just adds another actor -- the blockchain -- which can receive (but cannot send) messages and queries. 
Note that also in this case we must be very careful about security aspects.
In fact, often the hacks of dApps systems are made exploiting App System weaknesses, rather that SCs' ones.
\begin{enumerate}
\item Redefine the actors and the USs for the App System, starting from those gathered in steps \ref{item:actors} and \ref{item:USs}, adding the new actors represented by the SCs that interact with the App System. 
Define the acceptance tests of the App System.                            
\item Design the high-level architecture of the App System, including server and client tiers, and detail the way it accesses the blockchain.
The access can be done setting up and running one or more nodes of the blockchain, through an external provider, or using a standard wallet.
\item Define the UI of the App System, typically with a responsive approach, so that it can run on both mobile terminals and PCs. 
Having a fancy UI is of paramount importance to achieve the market success of the whole system. 
We suggest to perform UI design using well known standard approaches, such as Usage-Centered Design \cite{constantine1999} and Interaction Design \cite{sharp2019}. 
\item Define how the App System is decomposed in modules, their interfaces and the flow of messages between them. 
Define, if needed, the state diagrams of the modules, and the actions they take when events are raised by SCs. 
Define the structure and memorization of permanent data.
Select which data are anchored to the blockchain, by notarization of their hash digest.
Define the structure of the data or classes of the App System, including the flow of data and control between modules.
The interactions with the SCs must be consistent with the analysis of step 5.3. 
Since we use and agile approach, this design activity is not performed up-front, but through iterations that include coding and delivering increments of the App System, that is  implementation of the USs chosen for the iteration. 
As the App System is created, the need of additions or updates to the architecture may arise. 
Due to the above quoted security requirements, this design phase must be quite detailed, and made consistently with the corresponding activities of SCs design. 
UML class and sequence diagrams can help to design and document this system. 
\item 
Perform a security assessment of the external system, as described below in Sec. 4.3. 
\end{enumerate}

\item \textbf{Coding and testing the App System.} \label{item:code} 
In parallel to the SCs system, the App System is built and tested. 
Also in this case, we advocate following a Scrum approach. Alternatively, the team may chose a Lean-Kanban approach. Of course, the same approach should be used for both SCs and App System development. 
We stress that, if the developments of SCs and App System are made iteratively, every two or three iterations the results of the two branches must be integrated, as shown in Fig. \ref{fig:fig3}. If a continuous-flow, Lean-Kanban approach is performed, the integration should happen at the completion of every US, in both branches. The activities happening in parallel are:
\begin{enumerate}
\item Incrementally implement the USs of App System with an agile approach (Scrum or Kanban).
This step belongs to the "right flow" of ABCDE (see Fig.~\ref{fig:fig3}), and does not differ from the implementation of a Web application using Scrum or Kanban.
\item Perform the security assessment of the code written for the increment.
\item Write automated Unit Tests (UTs) and Acceptance Tests (ATs) for the USs implemented. Add the new tests to the test suite. Run the whole test suite to make sure that the additions did not break the system. 
\end{enumerate}

\item \textbf{Integrate, test and deploy the dApp system}. The integration of SCs with App System is performed every 2-3 development iterations of both systems.
\end{enumerate}

\subsection{UML diagrams for SCs} \label{subs:uml}
Nowadays, the most popular blockchain for dApp development is Ethereum, and the most used language is Solidity. 
This language is object-oriented (OOPL) because contracts are defined similarly to classes -- they have internal variables, and public and private functions. 
Each SC can inherit from one or more other contracts.
With respect to a standard OOPL, Solidity adds specific concepts like events and modifiers, and exhibits limitations in the types available for the SC data structure, and in the management of collections of data -- the only collections available so far are the array and the mapping. 
In the followings, we will describe an adaptation of UML diagrams specific for Solidity 0.5.
Possible modifications and extensions for other SC languages will be discussed in the section about future developments.

When designing and documenting SCs, graphic diagrams can be very useful to highlight the connections and the exchange of messages. To this purpose, we advocate the use of a subset of UML diagrams, being UML the universal standard for software design diagrams.
However, some specific concepts have to be introduced to account for peculiar SC features.
Luckily, UML has an extensibility mechanism called stereotype, which can be used to introduce new concepts, through tagging.

The UML diagrams we considered and modified to model SCs are Class diagrams, and Sequence diagrams. Also, UML Statecharts can be used to graphically represent the various states of a SC, or of a App System module and its transitions. 
Statecharts, however, do not need any specific stereotype.
We already suggested to use also the Use Case diagrams to model actors and related USs (in place of Use Cases).

The Class diagram enables to represent the structure and relationships of SCs. 
Table \ref{T1} shows the stereotypes we introduced in UML class diagrams in order to tag the SC specifities, and their description.

\begin{table*}[t] 
\centering
\caption{Additions to UML class diagram (stereotypes).}
\label{T1}
\begin{tabular}{|p{2.5 cm}|p{3.5 cm}|p{8 cm}|}
\hline
\multicolumn{1}{|l|}{\textbf{Stereotype}} & \textbf{Position} & \textbf{Description} \\ \hline
\guillemotleft contract\guillemotright &  Class symbol -- upper  compartment & Denotes a SC.  \\ \hline
\guillemotleft interface\guillemotright  & ditto & A kind of contract holding only function declarations \\ \hline
\guillemotleft library contract\guillemotright & ditto & A contract taken from a standard library \\ \hline
\guillemotleft enum\guillemotright  & ditto & A list of possible values, assigned to some variable. The values are listed in the middle compartment. The bottom compartment (holding operations) must be empty or absent.  \\ \hline
\guillemotleft struct\guillemotright & ditto & A record, defined in the data structure of a contract and used thereof, able to hold heterogeneous data. The fields are listed in the middle compartment. The bottom compartment must be empty or absent. \\ \hline
\guillemotleft event\guillemotright & Class symbol, middle compartment & An event that can be raised by the SC, signalling something relevant to external observers. \\ \hline
\guillemotleft modifier\guillemotright & Class symbol, bottom  compartment & A particular kind of guard function, called before another function \\ \hline
\guillemotleft array\guillemotright  & Class symbol, middle  compartment, or role of an association & A multiple variable, or 1:n relationship which is implemented using an array. \\ \hline
\guillemotleft mapping\guillemotright & ditto & The multiple variable, or 1:n relationship is implemented using a generic mapping. \\ \hline
\guillemotleft mapping [\textit{address}]\guillemotright  & ditto & A multiple variable, or 1:n relationship which is implemented using a mapping from an Ethereum address to the value. \\ \hline
\guillemotleft mapping [\textit{uint}]\guillemotright  & ditto & A multiple variable, or 1:n relationship which is implemented using a mapping from a unsigned integer to the value. \\ \hline
\end{tabular}
\end{table*}

In Solidity, there are no classes, but SCs are very similar to classes -- a SC has a data structure composed of variables, and functions able to access these variables.
Solidity source code can be used only for creating a SC. This is accomplished by using a special kind of transaction. The other two kinds of transactions are the transfer of Ethers, and the invocation of a function on an existing SC (message).
A piece of Solidity code can include several SCs, but a creation transaction can create at most one SC.
So, the other SCs can be used to be inherited from the created contract, or to specify the functions that are called on other existing SCs in the same blockchain, accessed through their address.
These relationships among SCs can be effectively captured by a UML class diagram.

To address the need to manage complex data, Solidity has the "struct" construct, similar to C, C++ and Java. In UML class diagram, we represent structs as classes, with a proper stereotype and with no bottom operation compartment.

A specific concept of Solidity are \textit{events}, raised when something relevant happens.
They can be caught by observer programs, able to act correspondingly. Remember that SCs cannot directly invoke functions of external systems.
Another peculiar concept of Solidity are the \textit{modifiers}.
These are boolean functions called before a function is executed.
They are able to check constraints, and possibly to stop the function execution. 

The last four stereotypes of Table \ref{T1} are about Solidity collections.
Owing to the limitations of blockchain storage, Solidity allows only two kinds of collections --  the array and the mapping. 
These stereotypes denote the kind of collection used for multiple variables of a data structure (middle compartment of UML class symbol), or for implementing an association, aggregation or composition. 
The array is an ordered set of values, indexed by their position, as in most computer languages. 
In Solidity, new values can be added to it, but not removed.
The corresponding stereotype is "<<array>>".

The mapping is able to store key-value pairs -- the keys being stored as hash values of the actual keys.
Given a key, a mapping can efficiently retrieve the value, but it is unable to iterate on its elements, both keys and values. 
Given the importance of the mapping in Solidity, we introduced three stereotypes to represent a mapping, denoted by the homonymous keyword. 
The first is the generic mapping; the second is the mapping having an Ethereum address as key, which is very used. The third refers to a common Solidity pattern -- using as keys positive, sequential integers, so that it is possible to iterate over them.

The other UML diagram very useful to represent the interactions among SCs and external actors is the Sequence Diagram.
These diagrams are used in UML to model messaging. 
In a blockchain, the relevant messages are related to the transactions, which in turn are sent from external actors, or from SCs to other SCs. 
Remember that messages are synonyms of "calls of public functions". 

A specific characteristic of Ethereum is that messages sent to a SC through a transaction take time (typically 15-20 seconds or more) to be answered.
However, if a message is sent to another SC during the execution of a function of a given SC, the time delay is negligible. 
This happens because the EVM, during the execution of the calling function, is able to locate in the blockchain and call any other SC.
To explicitly show this difference, which can be very important for security, GAS consumption and response time, we introduced the stereotypes <<trans-msg>> and <<direct-msg>> tagging the message calls sent through a transaction, and directly by a SC, respectively.
Note that the fact that UML Sequence Diagrams explicitly represent the flow of time from top to bottom of the diagram can also be used to quantify the timing difference between the two kinds of messages.

Another peculiarity of Ethereum is that a SC function which does not change the Blockchain is called a "view" function, and can be called immediately and at no cost. 
Again, this is because the EVM can locate the SC in the blockchain, verify that the function is "view" and call it very quickly and using a negligible amount of resources.
All other messages are executed only if proper GAS is paid.

Another kind of message that can be sent is the transfer of Ethers from an address to another. To represent this transfer, we use the Return Message of UML (a dashed arrow), tagged with the stereotype <<ethers>>.

Our Sequence Diagrams represent the message exchange among external actors and SCs, all called \textit{participants}, in a given scenario. 
The messages among external actors follow the usual UML notation. An external actors, however, can also send Ethers to another. 
The messages with at least a SC as sender or receiver belong to the following types:

\begin{itemize}
\item Transaction: characterized by coming from an external participant to a SC, it is validated and inserted in a block by miners.
\item Internal function call: a message sent by a SC to another SC that modifies the blockchain, thus costing GAS; it is represented as the usual "synchronous" or "asynchronous" message of UML Sequence diagrams.
\item SC creation: if sent by an external participant it is a transaction, if by another SC it is an internal call; in both cases it implies the call of the constructor of the new SC. In UML notation, creation is represented drawing the message arrow directly into the participant box representing the new SC.
\item View function call: a message to a SC which does not modify the blockchain, and costs no GAS; it is denoted by <<view>> or <<pure>> stereotype, and can be sent by both kinds of participant.
\item Fallback function call: the fallback function is a special function of each SC which is called whenever a function or an Ether transfer fail. This function implements recovery procedures, but can also be used to call whatever function of another contract, through the Proxy pattern. For this reason, we deem important to have a specific <<fallback>> stereotype.
\end{itemize}

Note that the above discussed characteristics are specific of Ethereum EVM, irrespectively of the specific high level language used to code the SCs.
Table \ref{T2} reports the stereotypes we introduced in UML Sequence diagrams to identify the participants sending messages (each having a unique address), and the kinds of messages they exchange. 

\begin{table}[ht]
\centering
\caption{The stereotypes added to UML Sequence diagrams.}
\label{T2}
\begin{tabular}{|c|p{2cm} | p{9.5cm}|}
\hline
\multicolumn{1}{|l|}{\textbf{Stereotype}} & \textbf{Position} & \textbf{Description} \\ \hline
\guillemotleft person\guillemotright & Participant box & A human role who posts transactions bearing messages, through wallet or some application. \\ \hline
\guillemotleft system\guillemotright & ditto & An external software system, able to send transactions to the Blockchain. \\ \hline
\guillemotleft device\guillemotright & ditto  & An IoT device, able to send transactions to the Blockchain. \\ \hline
\guillemotleft contract\guillemotright & ditto & A SC belonging to the system. \\ \hline
\guillemotleft external contract\guillemotright & ditto & A SC external to the system. \\ \hline
\guillemotleft oracle\guillemotright & ditto  & A particular type of SC, which holds information coming from the external world, provided by a trusted provider. \\ \hline
\guillemotleft account\guillemotright & ditto & An Ethereum address, just holding Ethers. It can only receive or send Ethers, when its owner activates the transfer. \\ \hline
\guillemotleft trans-msg\guillemotright & Message & The message is sent using an Ethereum transaction. \\ \hline
\guillemotleft direct-msg\guillemotright & Message & The message is sent by a SC, so it is executed immediately. \\ \hline
\guillemotleft view>> or <<pure\guillemotright & Message & The function called is of type "view" or "pure", so it costs no GAS. \\ \hline
\guillemotleft fallback\guillemotright & Message & Call to the fallback function. Only called by a SC on itself. \\ \hline
\guillemotleft ethers\guillemotright & Return Message & The dashed arrow represents a transfer of Ethers, and is can be shown also as a stand-alone message. \\ \hline
\end{tabular}
\end{table}

\subsection{Security assessment for Smart Contracts} \label{subs:security}
Assessing and defining patterns of good programming practice for Smart Contracts for granting security in dApps is still in its infancy and is an ongoing area of research. Nevertheless, based on the programmers' experience and on recent exploited weaknesses --very (in)famous and critical also for the amount of real money involved--, some major advices for security assessment in Smart Contracts have been identified and discussed among the the Solidity developers community.
In fact, Ethereum and Blockchain ecosystem are highly new and still somewhat experimental; in addition, SCs are often designed to handle and transfer significant amount of money (in cryptocurrency, but easily exchangeable to real money). 
Therefore, it is necessary that they correctly achieve their purposes, but it is also crucial that their execution is secure against attacks.

The critical issues regarding the safety of a dApp can be divided in three areas:
\begin{itemize}
\item \textit{Issues related to Blockchain itself}: the blockchain itself could be attacked. It is known, for instance, that blockchains using proof-of-work for block generation are subject, at least theoretically, to the so-called "51\% attack". Those based on proof-of-stake are vulnerable to other types of attack, for example to "fake stake attack". Using Ethereum technology, the use of the main net lowers the probability of a "51\% attack", given the number and the computing power fielded by the miners. Instead, using Ethereum Classic blockchain, a fork derived from Ethereum in 2016, the probability is higher because its miners' computer power is much lower. Using a permissioned blockchain, for instance Ethereum Parity "proof-of-authority", there is no "51\% attack", but the blockchain security depends on the honesty and reliability of the validating members, and on their control over their respective IT services. Clearly, this kind of attacks are are more a problem of design choice of the technology to be used than of proper dApp design, so their prevention go beyond the scope of this paper.

\item \textit{Issues related to SCs}: the most critical part of a dApp are the SCs, whose bytecode is publicly available, and exposed to all possible exploits. Moreover, developers often lack a full knowledge about implementation and usage of SCs, due to the the fact that this technology is in its early stage, it is evolving fast and is different from traditional development. In literature there are several analyses of possible vulnerabilities related to both Ethereum virtual machine and Solidity language \cite{huang2019} \cite{prait2019} \cite{liu2019}. These are a good starting point for providing a checklist of patterns to verify the SCs under development.

\item \textit{Issues related to the App System}: The App System is composed of the server and client side of the dApp, interacting with the SCs on one side, and with human roles and IoT devices and other systems on the other side. It must be designed and implemented with care, but it is somewhat less critical, provided that all best practices related to the security of Web applications are used; a special emphasis must be made to safeguard the access to the private keys of the various actors. We will not cover general Web security practices in this paper.

\end{itemize}

In the following of this section, we focus on security assurance practices regarding SC design and coding, which are the most critical and less studied among the issues cited above.

\subsubsection{General concepts of dApp security}
The first and foremost concept in security management is to have a security mindset. 
The development team(s), and the whole organization, must be fully aware of the importance of security and protection from attacks. 
Since ABCDE is an agile process, it is based on principles and practices such as: maximize communication, short iterations, refactoring, continuous testing, simplicity, intention-revaling code, use of simple tools. 
All these practices are also good for security, but Agile means incremental development where USs are continuously completed and tested. This greatly helps productivity, but might be at the expense of security. 

A good starting point to focus on security are the Top 10 Proactive Controls of OWASP organization \cite{owasp2018}. Those most relevant for dApp security, ordered by importance, are:
\begin{itemize}
\item C1: Define Security Requirements. This looks straightforward, but it is not. You must explicitly define the security requirements needed for your system. The requirements can be written as USs, or as non-functional features, and should have acceptance tests in the form of test cases to confirm these requirements have been implemented.
\item C2: Leverage Security Frameworks and Libraries. Don't write everything from scratch, but reuse software that is security-hardened, is coming from trusted sources and is maintained up to date.
\item C5: Validate All Inputs. This should be performed for user inputs on server-side, because client-side validation can be bypassed. Also, let the SC itself perform validation of key data sent to it through messages.
\item C6: Implement Digital Identity. In a dApp environment, digital identities are guaranteed by addresses and by the ownership of the relative private key, so this control is quite straightforward.
\item C7: Enforce Access Controls. SC can check access levels of addresses through a mapping, and act accordingly.
\item C8: Protect Data Everywhere. In particular, be aware that data stored in a SC are always accessible to read, independently of their visibility.
\item C10: Handle All Errors and Exceptions. It is known that even small mistakes in error handling, or forgetting to handle errors can lead to catastrophic failures in distributed systems. This is particularly true for SCs.
\end{itemize}

Specifically related to SC security are the general guidelines reported in \cite{bestPractices}, section: "General Philosophy", which complement OWASP ones. Here will just report a short description, and give the names of related security patterns, reported below this list: 
\begin{enumerate}
\item Prepare for failure. Be able to respond to errors, also in the context of SCs, which cannot be changed once deployed. This is related to patterns 'Emergency stop', 'Rate limit', 'Balance limit' and 'Proxy'
\item Rollout carefully. Try your best to catch and fix the bugs before the SC is fully released. Test contracts thoroughly, and add tests whenever new attack vectors are discovered.
\item Keep SCs simple. Complexity increases the risk of errors, so ensure that SCs and functions are small and modular, reuse SCs that are proven, prefer clarity to performance.
\item Keep up to date. Keep track of new security developments and upgrade to the latest version of any tool or library as quickly as possible.
\item Be aware of blockchain properties. While your previous programming experience is also applicable to SC programming, there are several pitfalls to be aware of.
\end{enumerate}

\subsubsection{Security in the design phase}
In the design phase, developers must be aware of, and use security patterns, as reported in references \cite{wohrer2018}, \cite{bart2017}, \cite{proxyy2019}, which we refer to. 
Table \ref{tab:secPat} shows the main security patterns.

\begin{center}
\begin{longtable}{|p{1cm}|p{3cm}|p{10cm}|p{1cm}|}
\caption{Main security patterns} \label{tab:secPat} \\

\hline \multicolumn{1}{|c|}{\textbf{ID}} & {\textbf{Name}} & \multicolumn{1}{c|}{\textbf{Description}} & \multicolumn{1}{c|}{\textbf{Ref.}} \\ \hline 
\endfirsthead

\hline
CEI & Check-effect-interaction & When performing a function in a SC: first, check all the preconditions, then apply the effects to the contract's state, and finally interact with other contracts. Never alter this sequence. & \cite{wohrer2018} \\  \hline
ES & Emergency stop, also known as "Circuit breaker" & Incorporate an emergency stop functionality into the SC that can be triggered by an authenticated party to disable sensitive functions. This is very useful in the case of major bug or security issue.& \cite{wohrer2018} \\  \hline
SB & Speed bump & Slow down contract sensitive tasks, so when malicious actions occur, the damage is limited and more time to counteract is available. For instance, limit the amount of money a user can withdraw per day, or impose a delay before withdrawals. & \cite{wohrer2018} \\  \hline
RL & Rate limit & Regulate how often a task can be executed within a period of time, to limit the number of messages sent to a SC, and thus its computational load. & \cite{wohrer2018} \\  \hline
MU & Mutex & A mutex is a mechanism to restrict concurrent access to a resource. Utilize it to hinder an external call from re-entering its caller function again. & \cite{wohrer2018} \\  \hline
BL & Balance limit & Limit the maximum amount of funds held within a SC. & \cite{wohrer2018} \\  \hline
GC & Guard Check & Ensure that all requirements on a SC state and on function inputs are met. Use properly \textit{assert()}, \textit{require()} and \textit{revert()} to check user inputs, SC state, invariants. & \cite{ethsec2019} \\  \hline
WF & Withdrawal from Contracts, also known as "Pull over Push" & When you need to send Ethers or tokens to an address, don't send them directly. Instead, authorize the address' owner to withdraw the funds, and let s/he perform the job. & \cite{solidity2019}, \cite{ethsec2019} \\  \hline
AU & Authorization & Restrict the execution of code according to the caller address. This is accomplished using mappings of addresses, and is typically checked using modifiers. & \cite{bart2017} \\  \hline
OR & Oracle & An oracle is a SC providing data from outside the blockchain, which are in turn fed to the oracle by a trusted source. Here the security risk lies in how actually the source can be trusted. & \cite{bart2017} \\  \hline
RN & Randomness & Not really a pattern, but some guidelines to simulate randomness in a deterministic environment like that of SCs. It is possible to query an Oracle, to use values not predictable \textit{a priori} as the hash of a block not yet created.  & \cite{bart2017} \\  \hline
TC & Time constraint & A time constraint specifies when an action is permitted, depending on the time registered in the block holding the transaction. It is used in Speed bump and Rate limit patterns. & \cite{bart2017} \\  \hline
TE & Termination & Used when the life of a SC has come to an end. This can be done by inserting ad-hoc code in the contract, or calling selfdestruct function. Usually, only the contract owner is authorized to terminate a contract. & \cite{bart2017} \\  \hline
MH & Math & A logic which computes some critical operations, protecting from overflows, underflows or other undesired characteristics of finite arithmetic. & \cite{bart2017} \\  \hline
PD & Proxy Delegate & Proxy patterns are a set of SCs working together to facilitate upgrading of SCs, despite their intrinsic immutability. A Proxy is used to refer to another SC, whose address can be changed. This approach also ensure that blockchain resources are used sparingly, thus saving GAS. & \cite{proxyy2019}, \cite{ethsec2019} \\  \hline

\end{longtable}
\end{center}

Our approach consists in using two security checklists, one to be performed during and after design and design upgrades, the other during coding phases. The aim is to verify that all security patterns and practices concerning known problems are applied. 
These practices are complementary to the agile practices reported in sections \ref{subs:agile} and \ref{subs:process}.
Depending on the size of the project and the number of SCs, the checklist can be unique for the system, or you may use a separate checklist for each SC subsystem.

Tables \ref{tab:desCheck} and \ref{tab:codCheck} present the security assurance practices we propose. they describe the checks to be performed, a short description of the vulnerability/vulnerabilities and how to avoid it/them, and one or more references to learn more about the problem. From these tables, it is easy to extract two checklists to be used to perform security assurance during the design and the coding of the SC system, respectively. 

\begin{center}
\begin{longtable}{|p{3cm}|p{9cm}|p{1.5cm}|p{2cm}|}
\caption{Security assurance checklist for the design phase} \label{tab:desCheck} \\

\hline \multicolumn{1}{|c|}{\textbf{To Check}} & \multicolumn{1}{c|}{\textbf{Description}} & \multicolumn{1}{c|}{\textbf{Ref.}} & {\textbf{ Related patterns}} \\ \hline 
\endfirsthead

\hline
\textit{Re-entrancy} & Functions that could be called recursively, before the first invocations is finished. This may cause destructive consequences. Ensure state committed before an external call. & \cite{atzei2017} \cite{bestPractices} & CEI, MU \\ \hline
\textit{Dependencies} & Use audited and trustworthy dependencies to existing SCs and ensure that newly written code is minimized by using libraries. &  \cite{bestPractices} & \\ \hline
\textit{Multiple Inheritance Caution} & Solidity uses the "C3 linearization". This means that when a contract is deployed, the compiler will linearize the inheritance from right to left. Multiple overrides of a function in complex inheritance hierarchies could potentially interact in tricky ways. & \cite{bestPractices} & \\ \hline
\textit{Include a fail-safe mechanism} & It is important to have some way to update the contract in the case some bugs will be discovered. For example, it is possible to have a contract forwarding calls and data to the latest version of the contract. & & ES, SB, RL, PD \\ \hline  
\textit{Limit the amount of ether} & If the code, the compiler or the platform has a bug, the funds stored in your smart contract may be lost, so limit the maximum amount. Check that all money transfers are performed through explicit withdrawals made by the beneficiary.   & & RL, BL, WF\\  \hline  
\textit{Be careful with randomness} & Random number generation in a deterministic system is very difficult. Do not rely on pseudo-randomness for important mechanisms. Current best solutions include hash-commit-reveal schemes (ie. one party generates a number, publishes its hash to "commit" to the value, and then reveals the value later) and RANDAO. & \cite{huang2019} section III-A-7 &  \\  \hline
\textit{Be careful with Timestamp} & Be aware that the timestamp of a block can be manipulated by a miner; all direct and indirect uses of timestamp should be analyzed and verified. If the scale of your time-dependent event can vary by 30 seconds and maintain integrity, it is safe to use a timestamp. This include thing like ending of auctions, registration periods, etc. Do not use the \textit{block.number} property as a timestamp. & \cite{prait2019} section IV-C & TC \\  \hline
\textit{Never assume that a contract has zero balance} & Be aware of coding an invariant that strictly checks the balance of a contract. An attacker can forcibly send ether to any account and this cannot be prevented. & \cite{bestPractices} & \\  \hline  
\textit{Transaction Ordering} & Miners have the power to alter the order of transactions arriving in short times. Inconsistent transactions' orders, with respect to the time of invocations, can cause race conditions.& \cite{prait2019} section IV-B & TC \\ \hline
\end{longtable}
\end{center}

\subsubsection{Security in the coding phase}

During coding, one major class of problems derives from ``external calls'', namely from functions which recur to others' SC code for completing their execution. 
In fact, a SC can call another SC, exploiting the execution of code contained in the latter contract. The pattern can be recursive, so the called SC can in turn perform an external call, and so on. As a consequence, external calls must be treated like calls to `untrusted' software. They should be avoided or minimized, because some malicious code could be introduced somewhere in a SC belonging to this path, and any external call represents a security risk. 
A typical risk of such contract interaction is ``reentrancy'', namely the called  contract can call back the calling function before the overall function execution has been completed. This pattern has been performed in the DAO attack. 
When it is not possible to avoid external calls, label all the potentially unsafe variables, functions and contracts interfaces as untrusted. Also, follow the "Check-effect-interaction" pattern.

Another important tool for SC security and error handling is the use of \textit{assert()}, \textit{require()} and \textit{revert()} guard functions. They are a very powerful security tool, and are the subject of security pattern "Guard Check" presented in Table \ref{tab:secPat}.
In general, use \textit{assert()} to check for invariants, to validate state after making changes, to prevent wrong conditions; if an \textit{assert()} statement fails, something very wrong happened and you need to fix the code.
Use \textit{require()} when you want to validate: user inputs, state conditions preceding an execution, or the response of an external call.
Use \textit{revert()} to handle the same type of cases as \textit{require()}, but with more complex logic \cite{bestPractices}.

The most important tool to achieve security and correctness, however, is to apply thorough, automated tests. This is even more crucial when writing SCs, because it is difficult or impossible to update a SC.
ABCDE does not prescribe the use ofspecific testing practices, such as Test Driven Design, but highlights the importance of testing. Presently, the most popular testing framework for Ethereum dApps is Truffle, whose website also provides documentation on how to test SCs and App System code -- see \cite{tru2019}, section: Testing Your Contracts.

The checklist for security assessment in the coding phase is reported in Table \ref{tab:codCheck}.

\begin{center}
\begin{longtable}{|p{3cm}|p{9cm}|p{1.5cm}|p{2cm}|}
\caption{Security assurance checklist for the coding phase} \label{tab:codCheck} \\

\hline \multicolumn{1}{|c|}{\textbf{To Check}} & \multicolumn{1}{c|}{\textbf{Description}} & \multicolumn{1}{c|}{\textbf{Ref.}} & {\textbf{ Related patterns}} \\ \hline 
\endfirsthead

\hline
\textit{External calls} & If possible, avoid them. When using low-level call functions (\textit{address.call(), callcode(), delegatecall()} and \textit{send()}) make sure to handle the possibility that the call will fail, by checking the return value. Also, avoid combining multiple ether transfers in a single transaction. Mark untrusted interactions: name the variables, methods, and contract interfaces of the functions that call external contracts, in a way that makes it clear that interacting with them is potentially unsafe.
& \cite{bestPractices} & CEI, MU, GC, WF \\ \hline
\textit{Prevent overflow and underflow} & If a balance reaches the maximum uint value it will circle back to zero; similarly, if a uint is made to be less than zero, it will cause an underflow and get set to its maximum value. One simply solution is to use a library like \textit{SafeMath.sol} by OpenZeppelin. & \cite{prait2019} section III-C & MH, GC \\ \hline
\textit{Beware of rounding errors} & All integer divisions round down to the nearest integer. Check that truncation does not produce unexpected behaviour (locked funds, incorrect results). & \cite{bestPractices} & MH, GC \\  \hline
\textit{Validate inputs to external and public functions} & Make sure the requirements are verified and check for arguments. & \cite{prait2019} section IV-F &  GC\\  \hline
\textit{Prevent unbounded loops} & The gas consumed increases with each iteration until it hits the block's gasLimit, stopping the execution. & \cite{ethsec2019} & \\  \hline 
\textit{tx.origin} & It is a global variable that returns the address of the message sender. Do not use \textit{tx.origin} as an authorization mechanism. & \cite{Mense:2018:SVE:3282373.3282419} section 3.1 &\\  \hline
\textit{Fallback functions} & Fallback functions are called when a contract receive a message without arguments and when no other function matches. You should keep them simple and check that the data is empty to avoid malicious invocation. & \cite{bestPractices} \cite{prait2019} section IV-A & CEI, MU, GC \\ \hline
\textit{Check if built-in variables or functions were overridden} & It is currently possible to override built-in globals in Solidity, such as . This allows SCs to override the functionality of built-ins such as \textit{msg} and \textit{revert()}. Although this is intended, it can mislead users of a SC, so the whole SC code must be checked. & \cite{bestPractices} & GC \\ \hline
\textit{Use interface type instead of the address for type safety} & When a function takes a contract address as an argument, it is better to pass an interface or contract type rather than raw \textit{address}. If the function is called elsewhere within the source code, the compiler it will provide additional type safety guarantees. & \cite{bestPractices} & GC \\ \hline
\textit{Enforce invariants with \textit{assert()}} & An assert guard triggers when an assertion fails - for instance an invariant property changing. You can verify it with a call to \textit{assert()}. Assert guards should be combined with other techniques, such as pausing the contract and allowing upgrades. (Otherwise, you may end up stuck, with an assertion that is always failing.) &  \cite{bestPractices} & GC \\  \hline
\textit{Lock pragmas to specific compiler version} & Contracts should be deployed with the same compiler version and flags that they have been tested with, so locking the version helps avoid the risk of undiscovered bugs. & \cite{bestPractices} & \\ \hline
\textit{Fix compiler warnings} & Take warnings seriously and fix them. Always use the latest version of the compiler to be notified about all recently introduced warnings. & \cite{solidity2019} & \\  \hline
\textit{Testing} & Be sure to have a $100\%$ text coverage and cover all critical edge cases with unit tests. Do not deploy recently written code, especially if it was written under tight deadline. & & \\  \hline 
\end{longtable}
\end{center}

\subsection{Gas optimization}

Besides security, another important factor of SCs that must be carefully designed since the beginning is their cost. 
Creating SCs and writing permanent data in a public blockchain can be very costly, so it is important to keep them to a minimum, and to limit the transactions that write or modify these data. 
Also, the messages exchanged among the App System and the SCs, and among SCs, must be properly designed and well documented.
Table \ref{tab:secPat} shows some specific patterns that can be used to save GAS.

Note that in Ethereum the maximum size of the bytecode of a SC is restricted to 24 KBytes by the standard EIP 170 (see section 13.4.2 of \cite{xu2019}). For serious SCs, that size limit can be hit easily, so many of the GAS saving patterns are useful also to make a SC viable.

\begin{center}
\begin{longtable}{|p{1.5cm}|p{3cm}|p{9cm}|p{1,5cm}|}
\caption{Main GAS saving patterns} \label{tab:secPat} \\

\hline \multicolumn{1}{|c|}{\textbf{ID}} & {\textbf{Name}} & \multicolumn{1}{c|}{\textbf{Description}} & \multicolumn{1}{c|}{\textbf{Ref.}} \\ \hline 
\endfirsthead

\hline
PD & Proxy Delegate & When you need to call external SCs, do not include their code. Include their interface and use the Proxy pattern, which uses the fallback function to call the SC functions. This is the same pattern also shown in Table \ref{tab:secPat} & \cite{proxyy2019}  \\  \hline
LS & Limit Storage & Limit data stored in the blockchain. Store non-permanent data in memory. Avoid changing storage data during computations -- change them only after all the calculations.  & \cite{gupta2018}  \\  \hline
PK & Pack your variables & In Ethereum, you pay GAS for every storage slot of 256 bits you use. You can pack as
many variables as you want in it, but you must order their declaration properly. Use integers smaller than 256 bits only if you have many to pack. If not, using 256 bits integers avoids the needed conversion to 256 bits, which costs GAS. Remember that elements in memory and call data are not packed. Use datatype \textit{bytes32} rather than \textit{bytes} or \textit{string}, if possible. Limit constant strings, for instance those used in \textit{require()} to explicit the error, to fit in 32 bytes. & \cite{gupta2018}  \\  \hline
DV & Delete variables no more needed & If you don’t need a variable anymore, delete it using the \textit{delete} keyword. In Ethereum, you get a GAS refund for freeing up storage space.  & \cite{gupta2018} \\  \hline
NI & Do not initialize variables with default values & All variables are initialized to zeroes at no cost. Do not explicitly initialize them to zero, or a value is given to them anyway when they are used.  & \cite{gupta2018} \\  \hline
MP & Use Mappings & To manage lists of data, use mappings with integer key and not arrays. This is known to save blockchain space. & \cite{gupta2018} \\  \hline
EP & Execution Paths & Thoroughly examine all possible execution paths, looking for code whose execution can be spared. Avoid repetitive checks of variables. Logical operators '||' and '\&\&' evaluate only the first operand if the second is not needed, so order them to maximize the probability that only one operand is computed. & \cite{gupta2018} \\  \hline
LE & Limit external calls & Limit calls to other SCs. Note that calling \textit{external} functions is cheaper than calling \textit{public} functions. The cheapest calls, however, are those to \textit{internal} functions.  & \cite{gupta2018}  \\  \hline
LM & Limit modifiers & . The code of modifiers is "inlined" inside the modified function, thus costing GAS. Internal functions, on the other hand, are not inlined but called as separate functions. They are very slightly more expensive in run time but save a lot of redundant bytecode in deployment, if used more than once. & \cite{gupta2018}  \\  \hline
UL & Use libraries & . The bytecode of external libraries is not made part of your SC, thus saving GAS. However, calling them is costly and has security issues. Use libraries for complex tasks. & \cite{gupta2018}  \\  \hline
EL & Event Log & If the App System needs to retrieve information about past events, that is not useful for SC execution, let the app directly access the Event Log in the blockchain. Note that if the event happened far in time, the time to retrieve it may be long. & \cite{gupta2018}  \\  \hline

\end{longtable}
\end{center}

\section{Experimental Validation}
\label{S:5}

The development process which later was named ABCDE was first devised in 2018 \cite{march2018}, and since then it has been used in several project carried on in our University group, and in firms we are consulting.
Among the projects which were developed, or which are in development, we may quote a system to trace the provenance of foods, a supply chain management system, a system to manage temporary job contracts, a voting system which was used to reward the best presentation at a conference, another one managing voting in firm shareholders' and board of directors meetings, a system to manage energy exchange in local networks of electricity producers and consumers, a system to automate agile software development \cite{march2019}.

The feedback of dApp developers using ABCDE method was generally positive, and was used to improve the method -- especially concerning security and GAS optimization practices.

Here we present, as an example of ABCDE usage, a simplified version of a dApp application aiming to implement a decentralized exchange (DEX) for tokens managed on Ethereum blockchain. 
A DEX is a system enabling the exchange of different tokens between two holders, who interact directly, without intermediaries. 
We started from the well-known 0x protocol project, the subject of a successful ICO held in 2017. The specification of the DEX can be found in the 0x Whitepaper \cite{warren2017}. 
We present a simplified version of the whole system. In particular, we dropped the part related to the protocol token (Section 4 of the Whitepaper).
Moreover, for the sake of brevity, we will not present the coding phases (phases 6, 8 and 9), but we stop at the end the design phases (phases 5 and 7).
The steps of ABCDE are presented below.

\begin{enumerate}
    \item \textbf{Goal of the system}. \textit{To manage a decentralized exchange, able to enable pairs of ERC20 and ERC721 token holders to exchange their tokens at an agreed rate on the Ethereum blockchain}.

\item \textbf{Actors}. The system has the following actors:
\begin{itemize}
    \item \textbf{Trader}: owner of tokens, wishing to post an offer, or to accept a posted offer.
    \item \textbf{Maker}: a trader who posts an offer to sell a given amount of her/his tokens, in exchange to tokens of another type, at a given exchange rate.
    \item \textbf{Taker}: a trader who accepts the offer of a Maker.
    \item \textbf{Relayer}: a system which facilitates signaling between market participants by hosting and propagating an order book of the offers.
    \item \textbf{DEX}: smart contract(s) on the Ethereum blockchain which accept orders signed by both a Maker and Taker, and activate the exchange of tokens.
    \item \textbf{Token}:  a SC on the Ethereum blockchain, managing a given token according to the ERC20 or ERC721 protocols.

\end{itemize}

\begin{figure}[ht]
\centering 
\includegraphics[width=14cm]{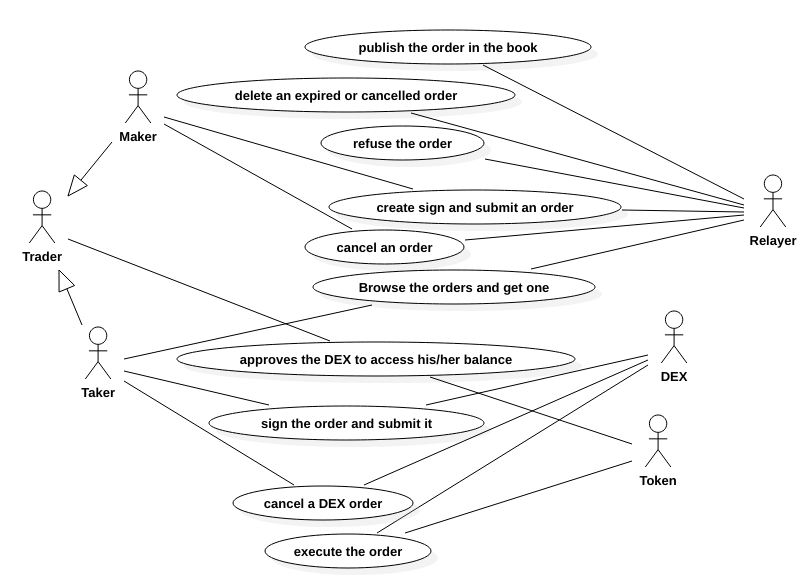}
\caption{The User Stories of the DEX system specification.}
\label{fig:fig4}
\end{figure}

\item \textbf{User Stories}. Fig. \ref{fig:fig4} shows the actors and the USs they are involved in, using a UML Use Case diagram, where the use cases are in fact USs. Note that these USs just specify the DEX, and do not depend on the specific technology used to implement it, except for the Ethereum blockchain, which the DEX necessarily has to interact with.
Here we have no room to show the USs in detail, so we refer the readers to the above quoted Whitepaper \cite{warren2017}. Instead, in Fig. \ref{fig:fig5} we show the UML class diagram derived by an analysis of the given USs. This diagram is not bound to a specific implementation of the relayer system, but just shows schematically the entities, the data structures and the operations emerging from the USs of Fig. \ref{fig:fig4}.

\begin{figure}[ht]
\centering 
\includegraphics[width=14cm]{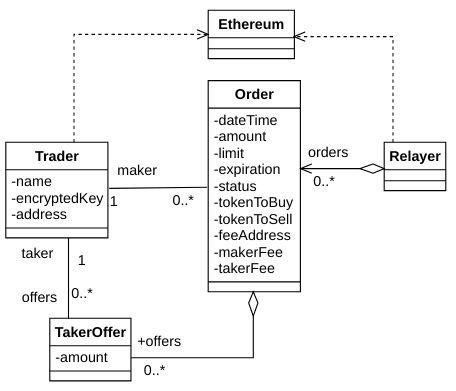}
\caption{The standard UML class diagram derived from the USs.}
\label{fig:fig5}
\end{figure}

\item \textbf{Divide the system into SC and App subsystems}. In this case the subdivision is trivial, because the Relayer system is a typical Web application, whereas the DEX and the Tokens are Smart Contracts by design.
The USs of the external app subsystem are the same of those reported in Fig. \ref{fig:fig4}, except the last one ("execute the order"), which is carried on solely by SCs.
As regards the blockchain subsystem, the US to implement are basically the messages to submit an order, or to cancel an order, sent by a Taker to the DEX. In practice, the actual implementation of the DEX contracts made by 0X Team is very complex, due to the strict security requirements, and to the many checks that must be made before performing the actual token transfer.

\begin{figure}[ht]
\centering 
\includegraphics[width=14cm]{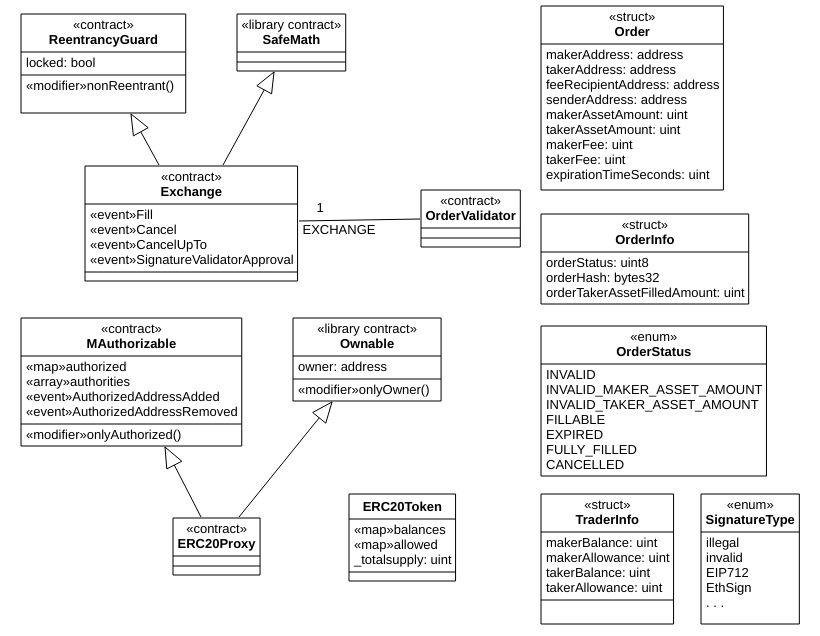}
\caption{The modified UML diagram, showing the structure of the required SCs of the DEX system.}
\label{fig:fig6}
\end{figure}

\item \textbf{Design of the SC subsystem}. The SC system is very complex, and a detailed description of its architecture is well beyond the scope of this paper. We report in Fig. \ref{fig:fig6} just a simplified UML class diagrams showing some of the actual SCs, to show some of the specific stereotypes used to document an SC system, as described in section \ref{subs:uml}.
The modifiers and the events enforcing the constraints relevant for the DEX are shown in Table \ref{T7} and Table \ref{T8}, respectively.

\begin{table*}[t] 
\centering
\caption{The \textit{modifiers} of the SCs.}
\label{T7}
\begin{tabular}{|p{3 cm}|p{10 cm}|}
\hline
\multicolumn{1}{|l|}{\textbf{Modifier}} & \textbf{Action -- Notes} \\ \hline
onlyOwner() & Enforces that the sender of the message is the owner of the contract. Inherited by Ownable standard contract
  \\ \hline
nonReentrant() & Enforces that the message is not recursively sent  \\ \hline
onlyAuthorized() & Enforces that the sender of the message belongs to a list of authorized addresses managed by the same SC
 \\ \hline
\end{tabular}
\end{table*}

\begin{table*}[t] 
\centering
\caption{The \textit{events} of the SCs.}
\label{T8}
\begin{tabular}{|p{5 cm}|p{9 cm}|}
\hline
\multicolumn{1}{|l|}{\textbf{Event}} & \textbf{Action -- Notes} \\ \hline
Fill & An order has been filled in the DEX. \\ \hline
Cancel & An order has been cancelled in the DEX. \\ \hline
CancelUpTo &  An order has been partially cancelled from the DEX. \\ \hline
AuthorizedAddressAdded & An address has been added to the list of authorized ones. \\ \hline
AuthorizedAddressRemoved & An address has been removed from the list of authorized ones. \\ \hline
\end{tabular}
\end{table*}

Fig. \ref{fig:fig7} shows a UML sequence diagram representing the interactions among most Actors of the systems, when a Taker accepts an order seen in the Relayer's book, and sends it to the DEX for execution, including the messages exchanged among the SCs.

\item  \textit{Omitted}.
\item \textbf{Design of App System}.  The App System is composed of the software able to present the present offers of tokens posted by the takers, and of the software used by takers and makers, respectively to post, modify or delete offers, and to accept offers. The latter software must be provided of a wallet able to store Ethers and send transactions to Ethereum blockchain. The design of this subsystem includes that of its user interfaces. The system is fairly complex, and the wallets must be designed and implemented using strong security practices. We will not dig further into this subsystem because, except for the wallet, it is a standard, Web-based system.
\item  \textit{Omitted}.
\item  \textit{Omitted}.
\end{enumerate}

\begin{figure}[ht]
\centering 
\includegraphics[width=16cm]{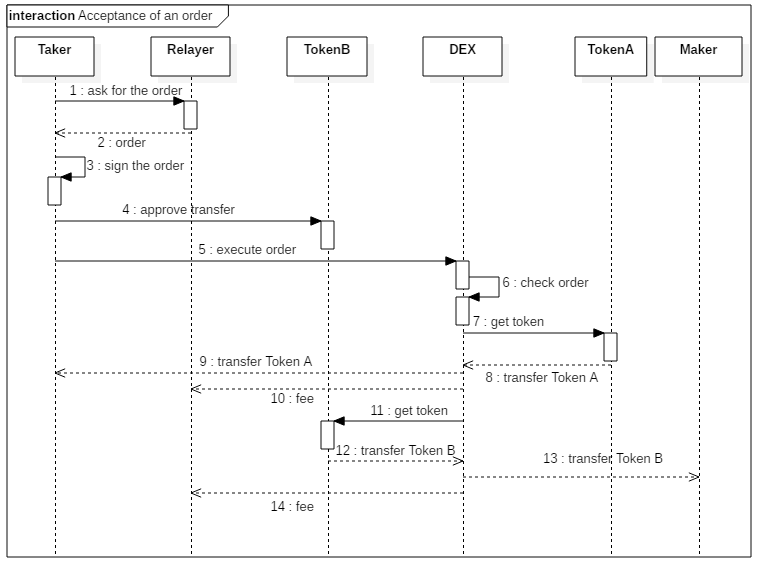}
\caption{The UML sequence diagram showing a Taker accepting an offer and sending it to the DEX for execution.}
\label{fig:fig7}
\end{figure}

\section{Conclusion and Future Work}
\label{S:6}

Despite the huge effort presently ongoing in developing dApps, software engineering practices are still poorly applied in software development of blockchain systems. 
The field is in fact still in its infancy, and tools or techniques for modeling and managing the peculiarities a software developer must face when dealing with blockchain-oriented software systems are still matter for researchers. 
Tools and techniques of traditional software engineering have not yet been adapted and modified to adhere to this new software paradigm. 
A sound software engineering approach might greatly help in overcoming many of the issues plaguing blockchain development providing developers with instruments similar to those typically used in traditional software engineering to afford architectural design, security issues, testing planes and strategies and to improve software quality and maintenance. 

Researchers in software engineering have a big opportunity to start studying a field that is very important and brand-new exploiting concepts, tools, instruments and ideas already consolidated in software engineering and changing and adapting them to this new software technology.

This work, whose a first version was presented in \cite{march2018}, moves toward this direction providing a full modeling of interactions among traditional software and blockchain environment, including Class diagrams, Statecharts, US’s diagrams, Sequence diagrams, Smart Contracts diagrams -- all specialized for blockchain application development.
It also provides a general scheme for managing blockchain development processes, and a simplified example of a Distributed Exchange Smart Contract, taken from a real set of SCs implementing a DEX. 

We believe that our work can be really valuable to blockchain firms, including ICO startups, that could develop a competitive advantage using SE (BOSE) practices since the beginning. 
The proposed method has also the potential to be applied to other SC environments, such as Hyperledger, Ripple and others, and we are exploring these extensions.

\section{ACKNOWLEDGMENTS}
This work was partially founded by the AIND project (Native Digital Administrations and Enterprises), funded by Sardinia Region, PIA call 2013, E.U. P.O. FESR 2007/2013, n. 3706 Rep. n. 316, 22/04/2016.

\bibliographystyle{model1-num-names}
\bibliography{biblio}

\begin{thebibliography}{55}
\expandafter\ifx\csname natexlab\endcsname\relax\def\natexlab#1{#1}\fi
\providecommand{\bibinfo}[2]{#2}
\ifx\xfnm\relax \def\xfnm[#1]{\unskip,\space#1}\fi
\bibitem[{Nakamoto(2008)}]{nakamoto2008}
\bibinfo{author}{S.~Nakamoto}, \bibinfo{title}{Bitcoin: A peer-to-peer
  electronic cash system}, \bibinfo{year}{2008}.
  \bibinfo{note}{Url="https://bitcoin.org/bitcoin.pdf", last accessed: August
  15, 2019}.
\bibitem[{Wood(2014)}]{wood2014}
\bibinfo{author}{G.~Wood}, \bibinfo{title}{Ethereum: A secure decentralised
  generalised transaction ledger}, \bibinfo{year}{2014}.
  \bibinfo{note}{Url="https://ethereum.github.io/yellowpaper/paper.pdf", last
  accessed (updated version): August 15, 2019}.
\bibitem[{Szabo(1997)}]{szabo1997}
\bibinfo{author}{N.~Szabo},
\newblock \bibinfo{title}{Smart contracts: Formalizing and securing
  relationships on public networks},
\newblock \bibinfo{journal}{First Monday} \bibinfo{volume}{2}
  (\bibinfo{year}{1997}).
  \bibinfo{note}{Url="https://ojphi.org/ojs/index.php/fm/article/view/548"}.
\bibitem[{Swan(2015)}]{swan2015}
\bibinfo{author}{M.~Swan}, \bibinfo{title}{Blockchain: Blueprint for a new
  economy}, \bibinfo{publisher}{O'Reilly Media, Inc.}, \bibinfo{year}{2015}.
\bibitem[{Biella and Zinetti(2016)}]{biella2016}
\bibinfo{author}{M.~Biella}, \bibinfo{author}{V.~Zinetti},
\newblock \bibinfo{title}{Blockchain technology and applications from a
  financial perspective},
\newblock \bibinfo{journal}{Unicredit Tehcnical Report}
  (\bibinfo{year}{2016}).
\bibitem[{Fenu et~al.(2018)Fenu, Marchesi, Marchesi, and Tonelli}]{fenu2018}
\bibinfo{author}{G.~Fenu}, \bibinfo{author}{L.~Marchesi},
  \bibinfo{author}{M.~Marchesi}, \bibinfo{author}{R.~Tonelli},
\newblock \bibinfo{title}{The ico phenomenon and its relationships with
  ethereum smart contract environment},
\newblock in: \bibinfo{booktitle}{Blockchain Oriented Software Engineering
  (IWBOSE), 2018 International Workshop on}, \bibinfo{organization}{IEEE}, pp.
  \bibinfo{pages}{26--32}.
\bibitem[{Chohan(2018)}]{chohan2018}
\bibinfo{author}{U.~Chohan}, \bibinfo{title}{The Problems of Cryptocurrency
  Thefts and Exchange Shutdowns}, \bibinfo{type}{Technical Report}, Discussion
  Paper Series: Notes on the 21 st Century, School of Business and Economics,
  University of New South Wales, Canberra, \bibinfo{year}{2018}.
\bibitem[{Atzei et~al.(2017)Atzei, Bartoletti, and Cimoli}]{atzei2017}
\bibinfo{author}{N.~Atzei}, \bibinfo{author}{M.~Bartoletti},
  \bibinfo{author}{T.~Cimoli},
\newblock \bibinfo{title}{A survey of attacks on ethereum smart contracts
  (sok)},
\newblock in: \bibinfo{editor}{M.~Maffei}, \bibinfo{editor}{M.~Ryan} (Eds.),
  \bibinfo{booktitle}{Principles of Security and Trust},
  \bibinfo{publisher}{Springer Berlin Heidelberg}, \bibinfo{year}{2017}, pp.
  \bibinfo{pages}{164--186}.
\bibitem[{Destefanis et~al.(2018)Destefanis, Marchesi, Ortu, Tonelli,
  Bracciali, and Hierons}]{destefanis2018}
\bibinfo{author}{G.~Destefanis}, \bibinfo{author}{M.~Marchesi},
  \bibinfo{author}{M.~Ortu}, \bibinfo{author}{R.~Tonelli},
  \bibinfo{author}{A.~Bracciali}, \bibinfo{author}{R.~Hierons},
\newblock \bibinfo{title}{Smart contracts vulnerabilities: A call for
  blockchain software engineering?},
\newblock in: \bibinfo{booktitle}{2018 International Workshop on Blockchain
  Oriented Software Engineering (IWBOSE)}.
\bibitem[{Porru et~al.(2017)Porru, Pinna, Marchesi, and Tonelli}]{porru2017}
\bibinfo{author}{S.~Porru}, \bibinfo{author}{A.~Pinna},
  \bibinfo{author}{M.~Marchesi}, \bibinfo{author}{R.~Tonelli},
\newblock \bibinfo{title}{Blockchain-oriented software engineering: challenges
  and new directions},
\newblock in: \bibinfo{booktitle}{Proceedings of the 39th International
  Conference on Software Engineering Companion}, \bibinfo{organization}{IEEE
  Press}, pp. \bibinfo{pages}{169--171}.
\bibitem[{Beck et~al.(2001)Beck, Beedle, Van~Bennekum, Cockburn, Cunningham,
  Fowler, Grenning, Highsmith, Hunt, Jeffries et~al.}]{beck2001}
\bibinfo{author}{K.~Beck}, \bibinfo{author}{M.~Beedle},
  \bibinfo{author}{A.~Van~Bennekum}, \bibinfo{author}{A.~Cockburn},
  \bibinfo{author}{W.~Cunningham}, \bibinfo{author}{M.~Fowler},
  \bibinfo{author}{J.~Grenning}, \bibinfo{author}{J.~Highsmith},
  \bibinfo{author}{A.~Hunt}, \bibinfo{author}{R.~Jeffries}, et~al.,
\newblock \bibinfo{title}{Manifesto for agile software development}
  (\bibinfo{year}{2001}).
\bibitem[{Chakraborty et~al.(2018)Chakraborty, Shahriyar, Iqbal, and
  Bosu}]{chakraborty2018}
\bibinfo{author}{P.~Chakraborty}, \bibinfo{author}{R.~Shahriyar},
  \bibinfo{author}{A.~Iqbal}, \bibinfo{author}{A.~Bosu},
\newblock \bibinfo{title}{Understanding the software development practices of
  blockchain projects: A survey},
\newblock in: \bibinfo{booktitle}{ESEM 2018, October 11–12, 2018, Oulu,
  Finland}, \bibinfo{organization}{ACM}.
\bibitem[{Bosu et~al.(2019)Bosu, Iqbal, Shahriyar, and Chakraborty}]{bosu2019}
\bibinfo{author}{A.~Bosu}, \bibinfo{author}{A.~Iqbal},
  \bibinfo{author}{R.~Shahriyar}, \bibinfo{author}{P.~Chakraborty},
\newblock \bibinfo{title}{Understanding the motivations, challenges and needs
  of blockchain software developers: a survey},
\newblock \bibinfo{journal}{Empirical Software Engineering}
  \bibinfo{volume}{24} (\bibinfo{year}{2019}) \bibinfo{pages}{2636--2673}.
\bibitem[{Schwaber and Beedle(2001)}]{schwaber2001}
\bibinfo{author}{K.~Schwaber}, \bibinfo{author}{M.~Beedle},
  \bibinfo{title}{Agile Software Development with Scrum},
  \bibinfo{publisher}{Pearson}, \bibinfo{year}{2001}.
\bibitem[{Beck(2002)}]{beck2002}
\bibinfo{author}{K.~Beck}, \bibinfo{title}{Test Driven Development: By
  Example}, \bibinfo{publisher}{Addison-Wesley Professional},
  \bibinfo{year}{2002}.
\bibitem[{Fowler(2018)}]{fowler2018}
\bibinfo{author}{M.~Fowler}, \bibinfo{title}{Refactoring: Improving the Design
  of Existing Code (2nd Edition)}, \bibinfo{publisher}{Addison-Wesley
  Professional}, \bibinfo{year}{2018}.
\bibitem[{Zheng et~al.(2018)Zheng, Xie, Dai, Chen, and Wang}]{zheng2018}
\bibinfo{author}{Z.~Zheng}, \bibinfo{author}{S.~Xie}, \bibinfo{author}{H.-N.
  Dai}, \bibinfo{author}{X.~Chen}, \bibinfo{author}{H.~Wang},
\newblock \bibinfo{title}{Blockchain challenges and opportunities: a survey},
\newblock \bibinfo{journal}{International Journal of Web and Grid Services}
  \bibinfo{volume}{14} (\bibinfo{year}{2018}) \bibinfo{pages}{352--375}.
\bibitem[{Tikhomirov(2017)}]{tikhomirov2017ethereum}
\bibinfo{author}{S.~Tikhomirov},
\newblock \bibinfo{title}{Ethereum: state of knowledge and research
  perspectives},
\newblock in: \bibinfo{booktitle}{International Symposium on Foundations and
  Practice of Security}, \bibinfo{organization}{Springer}, pp.
  \bibinfo{pages}{206--221}.
\bibitem[{cmc(2019)}]{cmc2019}
\bibinfo{title}{Coinmarketcap website}, \bibinfo{year}{2019}.
  \bibinfo{note}{Url="https://coinmarketcap.com/tokens/", last accessed: August
  12, 2019}.
\bibitem[{sod(2019)}]{sod2019}
\bibinfo{title}{State of the dapps website}, \bibinfo{year}{2019}.
  \bibinfo{note}{Url="https://www.stateofthedapps.com/stats", last accessed:
  November 16, 2019}.
\bibitem[{Dannen(2017)}]{dannen2017introducing}
\bibinfo{author}{C.~Dannen}, \bibinfo{title}{Introducing Ethereum and
  Solidity}, \bibinfo{publisher}{Springer}, \bibinfo{year}{2017}.
\bibitem[{Cohn(2004)}]{cohn2004}
\bibinfo{author}{M.~Cohn}, \bibinfo{title}{User Stories Applied: For Agile
  Software Development}, \bibinfo{publisher}{Addison-Wesley Professional},
  \bibinfo{year}{2004}.
\bibitem[{Janzen and Saiedian(2005)}]{janzen2005}
\bibinfo{author}{D.~Janzen}, \bibinfo{author}{H.~Saiedian},
\newblock \bibinfo{title}{Test-driven development concepts, taxonomy, and
  future direction},
\newblock \bibinfo{journal}{Computer} \bibinfo{volume}{38}
  (\bibinfo{year}{2005}) \bibinfo{pages}{43--50}.
\bibitem[{tru(2019)}]{tru2019}
\bibinfo{title}{Truffle website}, \bibinfo{year}{2019}.
  \bibinfo{note}{Url="https://www.trufflesuite.com/", last accessed: October 6,
  2019}.
\bibitem[{Beck(2000)}]{beck2000}
\bibinfo{author}{K.~Beck}, \bibinfo{title}{Extreme programming explained:
  embrace change}, \bibinfo{publisher}{Addison-Wesley professional},
  \bibinfo{year}{2000}.
\bibitem[{Rumbaugh et~al.(2017)Rumbaugh, Booch, and Jacobson}]{rumbaugh2017}
\bibinfo{author}{J.~Rumbaugh}, \bibinfo{author}{G.~Booch},
  \bibinfo{author}{I.~Jacobson}, \bibinfo{title}{The unified modeling language
  reference manual}, \bibinfo{publisher}{Addison Wesley}, \bibinfo{year}{2017}.
\bibitem[{Xu et~al.(2019)Xu, Weber, and Staples}]{xu2019}
\bibinfo{author}{X.~Xu}, \bibinfo{author}{I.~Weber},
  \bibinfo{author}{M.~Staples}, \bibinfo{title}{Architecture for Blockchain
  Applications}, \bibinfo{publisher}{Springer}, \bibinfo{year}{2019}.
\bibitem[{Xu et~al.(2017)Xu, Weber, Staples, Zhu, Bosch, Bass, Pautasso, and
  Rimba}]{xu2017}
\bibinfo{author}{X.~Xu}, \bibinfo{author}{I.~Weber},
  \bibinfo{author}{M.~Staples}, \bibinfo{author}{L.~Zhu},
  \bibinfo{author}{J.~Bosch}, \bibinfo{author}{L.~Bass},
  \bibinfo{author}{C.~Pautasso}, \bibinfo{author}{P.~Rimba},
\newblock \bibinfo{title}{A taxonomy of blockchain-based systems for
  architecture design},
\newblock in: \bibinfo{booktitle}{Software Architecture (ICSA), 2017 IEEE
  International Conference on}, \bibinfo{organization}{IEEE}, pp.
  \bibinfo{pages}{243--252}.
\bibitem[{Wessling et~al.(2018)Wessling, Ehmke, Hesenius, and
  Gruhn}]{Wessling2018}
\bibinfo{author}{F.~Wessling}, \bibinfo{author}{C.~Ehmke},
  \bibinfo{author}{M.~Hesenius}, \bibinfo{author}{V.~Gruhn},
\newblock \bibinfo{title}{How much blockchain do you need? towards a concept
  for building hybrid dapp architectures},
\newblock in: \bibinfo{booktitle}{WETSEB 2018-1st International Workshop on
  Emerging Trends in Software Engineering for Blockchain}.
\bibitem[{Fridgen et~al.(2018)Fridgen, Lockl, Radszuwill, Rieger, Schweizer,
  and Urbach}]{fridgen2018}
\bibinfo{author}{G.~Fridgen}, \bibinfo{author}{J.~Lockl},
  \bibinfo{author}{S.~Radszuwill}, \bibinfo{author}{A.~Rieger},
  \bibinfo{author}{A.~Schweizer}, \bibinfo{author}{N.~Urbach},
\newblock \bibinfo{title}{A solution in search of a problem: A method for the
  development of blockchain use cases},
\newblock in: \bibinfo{booktitle}{24th Americas Conference on Information
  Systems (AMCIS), New Orleans, USA, August 2018}.
\bibitem[{Jurgelaitis et~al.(2019)Jurgelaitis, Drungilas, Ceponiene, Butkiene,
  and Vaiciukynas}]{jurg2019}
\bibinfo{author}{M.~Jurgelaitis}, \bibinfo{author}{V.~Drungilas},
  \bibinfo{author}{L.~Ceponiene}, \bibinfo{author}{R.~Butkiene},
  \bibinfo{author}{E.~Vaiciukynas},
\newblock \bibinfo{title}{Modelling principles for blockchain-based
  implementation of business or scientific processes},
\newblock in: \bibinfo{booktitle}{Proceedings of the International Conference
  on Information Technologies}, IVUS 2019, pp. \bibinfo{pages}{43--47}.
\bibitem[{Beller and Hejderup(2019)}]{beller2019}
\bibinfo{author}{M.~Beller}, \bibinfo{author}{J.~Hejderup},
\newblock \bibinfo{title}{Blockchain-based software engineering},
\newblock in: \bibinfo{booktitle}{Proceedings of the 41th International
  Conference on Software Engineering Companion}, \bibinfo{organization}{IEEE
  Press}, pp. \bibinfo{pages}{53--56}.
\bibitem[{Lenarduzzi et~al.(2018)Lenarduzzi, Lonesu, Marchesi, and
  Tonelli}]{lenarduzzi2018}
\bibinfo{author}{V.~Lenarduzzi}, \bibinfo{author}{I.~Lonesu},
  \bibinfo{author}{M.~Marchesi}, \bibinfo{author}{R.~Tonelli},
\newblock \bibinfo{title}{Blockchain applications for agile methodologies},
\newblock in: \bibinfo{booktitle}{Proceedings of the 19th International
  Conference on Agile Software Development: Companion}, XP 2018,
  \bibinfo{publisher}{ACM}, \bibinfo{address}{New York, NY, USA},
  \bibinfo{year}{2018}, pp. \bibinfo{pages}{30:1--30:3}.
\bibitem[{Marchesi et~al.(2019)Marchesi, Destefanis, Lenarduzzi, Lunesu, Ortu,
  Pinna, and Tonelli}]{march2019}
\bibinfo{author}{M.~Marchesi}, \bibinfo{author}{G.~Destefanis},
  \bibinfo{author}{V.~Lenarduzzi}, \bibinfo{author}{M.~Lunesu},
  \bibinfo{author}{M.~Ortu}, \bibinfo{author}{A.~Pinna},
  \bibinfo{author}{R.~Tonelli},
\newblock \bibinfo{title}{Agile software development automated by blockchain
  smart contracts},
\newblock in: \bibinfo{booktitle}{Proceedings of the Software Engineering
  Conference Russia}, SECR 2019, \bibinfo{publisher}{ACM},
  \bibinfo{address}{New York, NY, USA}, \bibinfo{year}{2019}.
\bibitem[{Praitheeshan et~al.(2019)Praitheeshan, Pan, Yu, Liu, and
  Doss}]{prait2019}
\bibinfo{author}{P.~Praitheeshan}, \bibinfo{author}{L.~Pan},
  \bibinfo{author}{J.~Yu}, \bibinfo{author}{J.~Liu}, \bibinfo{author}{R.~Doss},
\newblock \bibinfo{title}{Security analysis methods on ethereum smart contract
  vulnerabilities: A survey},
\newblock \bibinfo{journal}{arXiv preprint arXiv:1908.08605}
  (\bibinfo{year}{2019}).
\bibitem[{Huang et~al.(2019)Huang, Bian, Li, Zhao, and Shi}]{huang2019}
\bibinfo{author}{Y.~Huang}, \bibinfo{author}{Y.~Bian}, \bibinfo{author}{R.~Li},
  \bibinfo{author}{L.~Zhao}, \bibinfo{author}{P.~Shi},
\newblock \bibinfo{title}{Smart contract security: A software lifecycle
  perspective},
\newblock \bibinfo{journal}{IEEE Access,} \bibinfo{volume}{7}
  (\bibinfo{year}{2019}).
\bibitem[{Baumeister et~al.(1999)Baumeister, Koch, and Mandel}]{baumeister1999}
\bibinfo{author}{H.~Baumeister}, \bibinfo{author}{N.~Koch},
  \bibinfo{author}{L.~Mandel},
\newblock \bibinfo{title}{Towards a uml extension for hypermedia design},
\newblock in: \bibinfo{booktitle}{International Conference on the Unified
  Modeling Language}, \bibinfo{organization}{Springer}, pp.
  \bibinfo{pages}{614--629}.
\bibitem[{Baresi et~al.(2001)Baresi, Garzotto, and Paolini}]{baresi2001}
\bibinfo{author}{L.~Baresi}, \bibinfo{author}{F.~Garzotto},
  \bibinfo{author}{P.~Paolini},
\newblock \bibinfo{title}{Extending uml for modeling web applications},
\newblock in: \bibinfo{booktitle}{System Sciences, 2001. Proceedings of the
  34th Annual Hawaii International Conference on},
  \bibinfo{organization}{IEEE}, pp. \bibinfo{pages}{10--pp}.
\bibitem[{Rocha and Ducasse(2018)}]{rocha2018preliminary}
\bibinfo{author}{H.~Rocha}, \bibinfo{author}{S.~Ducasse},
\newblock \bibinfo{title}{Preliminary steps towards modeling blockchain
  oriented software},
\newblock in: \bibinfo{booktitle}{WETSEB 2018-1st International Workshop on
  Emerging Trends in Software Engineering for Blockchain}.
\bibitem[{Coad et~al.(1991)Coad, Yourdon, and Coad}]{coad1991}
\bibinfo{author}{P.~Coad}, \bibinfo{author}{E.~Yourdon},
  \bibinfo{author}{P.~Coad}, \bibinfo{title}{Object-oriented analysis},
  volume~\bibinfo{volume}{2}, \bibinfo{publisher}{Yourdon press Englewood
  Cliffs, NJ}, \bibinfo{year}{1991}.
\bibitem[{Anderson(2010)}]{anderson2010}
\bibinfo{author}{D.~J. Anderson}, \bibinfo{title}{Kanban: successful
  evolutionary change for your technology business}, \bibinfo{publisher}{Blue
  Hole Press}, \bibinfo{year}{2010}.
\bibitem[{Constantine and Lockwood(1999)}]{constantine1999}
\bibinfo{author}{L.~L. Constantine}, \bibinfo{author}{L.~A. Lockwood},
  \bibinfo{title}{Software for use: a practical guide to the models and methods
  of usage-centered design}, \bibinfo{publisher}{Pearson Education},
  \bibinfo{year}{1999}.
\bibitem[{Sharp et~al.(2019)Sharp, Rogers, and Preece}]{sharp2019}
\bibinfo{author}{H.~Sharp}, \bibinfo{author}{Y.~Rogers},
  \bibinfo{author}{J.~Preece}, \bibinfo{title}{Interaction design: beyond
  human-computer interaction, 5th edition}, \bibinfo{publisher}{John Wiley \&
  Sons}, \bibinfo{year}{2019}.
\bibitem[{Liu and Liu(2019)}]{liu2019}
\bibinfo{author}{J.~Liu}, \bibinfo{author}{Z.~Liu},
\newblock \bibinfo{title}{A survey on security verification of blockchain smart
  contracts},
\newblock \bibinfo{journal}{IEEE Access,} \bibinfo{volume}{7}
  (\bibinfo{year}{2019}).
\bibitem[{Anton et~al.(2018)Anton, Manico, and Bird}]{owasp2018}
\bibinfo{author}{K.~Anton}, \bibinfo{author}{J.~Manico},
  \bibinfo{author}{J.~Bird}, \bibinfo{title}{OWASP Proactive Controls for
  Developers}, \bibinfo{type}{Technical Report}, Open Web Application Security
  Project (OWASP), \bibinfo{year}{2018}.
\bibitem[{bes(2019)}]{bestPractices}
\bibinfo{title}{Consensys solidity best practices website},
  \bibinfo{year}{2019}.
  \bibinfo{note}{Url="https://consensys.github.io/smart-contract-best-practices/",
  last accessed: November, 2019}.
\bibitem[{Wohrer and Zdun(2018)}]{wohrer2018}
\bibinfo{author}{M.~Wohrer}, \bibinfo{author}{U.~Zdun},
\newblock \bibinfo{title}{Smart contracts: Security patterns in the ethereum
  ecosystem and solidity},
\newblock in: \bibinfo{booktitle}{Blockchain Oriented Software Engineering
  (IWBOSE), 2018 International Workshop on}, \bibinfo{organization}{IEEE}, pp.
  \bibinfo{pages}{2--8}.
\bibitem[{Bartoletti and Pompianu(2017)}]{bart2017}
\bibinfo{author}{M.~Bartoletti}, \bibinfo{author}{L.~Pompianu},
\newblock \bibinfo{title}{An empirical analysis of smart contracts: platforms,
  applications, and design patterns},
\newblock in: \bibinfo{editor}{B.~M. et~al.} (Ed.),
  \bibinfo{booktitle}{Financial Cryptography and Data Security. FC 2017.
  Lecture Notes in Computer Science}, \bibinfo{publisher}{Springer, Cham},
  \bibinfo{year}{2017}, pp. \bibinfo{pages}{494--509}.
\bibitem[{pro(2019)}]{proxyy2019}
\bibinfo{title}{Proxy patterns}, \bibinfo{year}{2019}.
  \bibinfo{note}{Url="https://blog.openzeppelin.com/proxy-patterns/", last
  accessed: November, 2019}.
\bibitem[{eth(2019)}]{ethsec2019}
\bibinfo{title}{Ethereum smart contract security best practices website},
  \bibinfo{year}{2019}.
  \bibinfo{note}{Url="https://ethereum-contract-security-techniques-and-tips.readthedocs.io/en/latest/",
  last accessed: November, 2019}.
\bibitem[{sol(2019)}]{solidity2019}
\bibinfo{title}{Solidity website}, \bibinfo{year}{2019}.
  \bibinfo{note}{Url="https://solidity.readthedocs.io/en/v0.5.13/index.html",
  last accessed: November, 2019}.
\bibitem[{Mense and Flatscher(2018)}]{Mense:2018:SVE:3282373.3282419}
\bibinfo{author}{A.~Mense}, \bibinfo{author}{M.~Flatscher},
\newblock \bibinfo{title}{Security vulnerabilities in ethereum smart
  contracts},
\newblock in: \bibinfo{booktitle}{Proceedings of the 20th International
  Conference on Information Integration and Web-based Applications \&
  Services}, iiWAS2018, \bibinfo{publisher}{ACM}, \bibinfo{address}{New York,
  NY, USA}, \bibinfo{year}{2018}, pp. \bibinfo{pages}{375--380}.
\bibitem[{Gupta(2018)}]{gupta2018}
\bibinfo{author}{M.~Gupta}, \bibinfo{title}{Solidity gas optimization tips},
  \bibinfo{year}{2018}.
  \bibinfo{note}{Url="https://mudit.blog/solidity-gas-optimization-tips/", last
  accessed: November, 2019}.
\bibitem[{Marchesi et~al.(2018)Marchesi, Marchesi, and Tonelli}]{march2018}
\bibinfo{author}{M.~Marchesi}, \bibinfo{author}{M.~Marchesi},
  \bibinfo{author}{R.~Tonelli},
\newblock \bibinfo{title}{An agile software engineering method to design
  blockchain applications},
\newblock in: \bibinfo{booktitle}{Proceedings of the Software Engineering
  Conference Russia}, SECR 2018, \bibinfo{publisher}{ACM},
  \bibinfo{address}{New York, NY, USA}, \bibinfo{year}{2018}.
\bibitem[{Warren and Bandeali(2017)}]{warren2017}
\bibinfo{author}{W.~Warren}, \bibinfo{author}{A.~Bandeali}, \bibinfo{title}{0x
  An open protocol for decentralized exchange on the ethereum blockchain},
  \bibinfo{year}{2017}.
  \bibinfo{note}{Url="https://0xproject.com/pdfs/0x\_white\_paper.pdf", last
  accessed: November, 2019}.

\end{thebibliography}











\end{document}